\documentclass[english,aps,prl, twocolumn,superscriptaddress,longbibliography]{revtex4-2}
\usepackage[utf8]{inputenc}
\usepackage{amsmath}
\usepackage{amssymb}
\usepackage{bm}
\usepackage{graphicx}
\usepackage{hyperref}
\usepackage{braket}
\usepackage{babel}
\usepackage{mathtools}
\usepackage{xspace}
\usepackage{xcolor}
\usepackage{physics}
\usepackage{bbm}
\usepackage{times}
\usepackage{ulem}

\usepackage{xcolor}
\definecolor{darkblue}{HTML}{004D6B}
\definecolor{darkred}{HTML}{8c1515}
\definecolor{darkgreen}{HTML}{006400}
\usepackage{hyperref}
\hypersetup{
    pdftitle={NishimoriCatQubitLoss},
	colorlinks=true,
	urlcolor=darkred,
	citecolor=darkblue,
	linkcolor=darkred,
	breaklinks
}
\pagecolor{white}

\begin{document}
%%%%%%%%%%%%%%%%%%%%%%%%%%%%%%%%%%%%%%%%%%%%%%%%%%%%%%%%%%%%%%%%%%%
\title{Dynamical self-dual criticality in Fibonacci-monitored quantum Ising chains}

%%%%%%%%%%% Authors %%%%%%%%%%%%%%
\author{Finn Eckstein}
\thanks{These authors contributed equally to this work.}
\affiliation{Institute for Theoretical Physics, University of Cologne, Zulpicher Straße 77, 50937 Cologne, Germany}

\author{Harald Schmid}
\thanks{These authors contributed equally to this work.}
\affiliation{Technical University of Munich, TUM School of Natural Sciences, Physics Department, 85748 Garching, Germany}

\affiliation{Munich Center for Quantum Science and Technology (MCQST), Schellingstr. 4, 80799 M{\"u}nchen, Germany}

\author{Quinten Preiss}
\affiliation{Institute for Theoretical Physics, University of Cologne, Zulpicher Straße 77, 50937 Cologne, Germany}

\author{Simon Trebst}
\affiliation{Institute for Theoretical Physics, University of Cologne, Zulpicher Straße 77, 50937 Cologne, Germany}

\author{Felix von Oppen}
\affiliation{Dahlem Center for Complex Quantum Systems, Fachbereich Physik, and Halle-Berlin-Regensburg \\
Cluster of Excellence CCE, Freie Universit{\"a}t Berlin, 14195 Berlin, Germany}

\author{Guo-Yi Zhu}
\affiliation{The Hong Kong University of Science and Technology (Guangzhou), Nansha, Guangzhou, 511400, Guangdong, China}

%%%%%%%%%%%%%%%%%%%%%%%%%%%%%%%%%%%%%%%%%%%%%%%%%%%%%%%%%%%%%%%%%%%
\begin{abstract}
     For the quantum phase transition in the transverse-field Ising chain, Kramers-Wannier duality not only protects its critical properties but also pinpoints the location of the phase transition.    
     Its role in out-of-equilibrium, monitored dynamics, however, remains largely unexplored beyond time-periodic Floquet protocols 
     where self-duality turns into a statistical {\it average} symmetry.  
     Here we explore the emergence of {\it dynamical} self-duality in the absence of time-translation symmetry by investigating the
     monitored dynamics of one-dimensional Ising/Majorana chains where measurements are arranged in a {\it quasiperiodic} Fibonacci sequence.
     We find that the dynamical extension of this non-invertible symmetry to an out-of-equilibrium setting allows one to organize the dynamical 
     phase diagram of entangled phases, both predicting the transition locations and protecting universal critical behavior. 
     Analytically and numerically, we identify two distinct critical lines, both related to the golden ratio, for Born-rule weak measurements 
     and for random Clifford projective measurements.
     The latter coincides with the transition of a pure imaginary-time evolution, which can be viewed as a post-selected trajectory. 
     The universality classes of the long-time critical steady states at Fibonacci times are determined, while the transient dynamics between Fibonacci times is deformed by measurements, realizing {\it dynamical} measurement-altered quantum criticality in real time.  
     \end{abstract}

\date{\today}
\maketitle
%%%%%%%%%%%%%%%%%%%%%%%%%%%%%%%%%%%%%%%%%%%%%%%%%%%%%%%%%%%%%%%%%%%

%%%%%%%%%%%%%%%%%%%%%%%%%%%%%%%%%%%%%%%%%%%%%%%%%%%%%%%%%%%%%%%%%%%
% Introduction
%%%%%%%%%%%%%%%%%%%%%%%%%%%%%%%%%%%%%%%%%%%%%%%%%%%%%%%%%%%%%%%%%%%
\textit{Introduction.---} 
Quantum circuits enable unprecedented control over the entanglement dynamics of quantum many-body systems through discrete sequences of elementary gates.
With mid-circuit measurements, the resulting non-equilibrium {\it monitored dynamics}~\cite{Fisher2022reviewMIPT,Potter21review} is non-unitary and conceptually distinct from equilibrium Hamiltonian or driven Floquet dynamics.
A paradigmatic example is the monitored Ising chain where an observer monitors the Ising interaction and the transverse field
(via two-qubit parity checks and single-qubit measurements, respectively)~\cite{Buechler20, Hsieh2021measure}.
In contrast to its Hamiltonian variant~\cite{Pfeuty1970,Sachdev2011}, 
where the ground state undergoes an Ising transition between ferromagnetic and paramagnetic phases, 
the monitored circuit dynamics  
is best described in terms of a mixed state~\cite{Wang25selfdual}
undergoing a strong-to-weak spontaneous symmetry breaking (SW-SSB) transition~\cite{Wang24strtowksym,Luo24weaksym,You24weaksym,Sun2025,Ziereis2025}.
The critical state is drastically changed, from unitary Ising criticality to a weak self-dual, critical state~\cite{Wang25selfdual} described by a non-unitary conformal field theory (CFT)~\cite{Ludwig2020,Bao20learning,Pixley20mipt}. 
In the limit of projective measurements~\cite{Hsieh2021measure}, also dubbed the ``projective Ising model" \cite{Buechler20, Roser2023,Roser2026}, the dynamics reduces to a Clifford circuit whose criticality is described by the percolation CFT instead~\cite{Buechler20,Vedika2021measure,Barkeshli2021measure,Nahum2021forcedmeasurement,Zhu23structuredVolumeLaw,klocke2024entanglement}. 
The three distinct transitions -- Ising, weak self-dual, and percolation -- share that they obey Kramers-Wannier self-duality. While this is a strong symmetry for the Ising transition, it turns into a weak symmetry in the latter two cases, i.e.\ only the statistical ensemble
of trajectories exhibits Kramers-Wannier symmetry (but not individual trajectories).

When viewing the monitored Ising chain as a monitored Kitaev chain~\cite{Wang25selfdual, Nahum20freefermion, Graham23majorana,Kitaev2001} with a transition between topological and trivial phases, the self-duality at criticality simply reflects the spatial translation symmetry of the Majorana lattice, both for the Hamiltonian and the circuit model. In the latter case, a duality transform modifies the space-time circuit of measurements. In the absence of {\it time-translation symmetry},  
the monitored quantum system is no longer guaranteed to possess a self dual structure. In such a {\it dynamical} and {\it statistical} setting, the emergence of self-duality becomes highly non-trivial.

%%%%%%%%%%%%%%%%%%%%%%%%%%%%%%%%%%%%%%%%%%%%%%%%%%%%%%%%%%%%%%%%%%%
\begin{figure*}[t!]
    \centering
    \includegraphics[width=\textwidth]{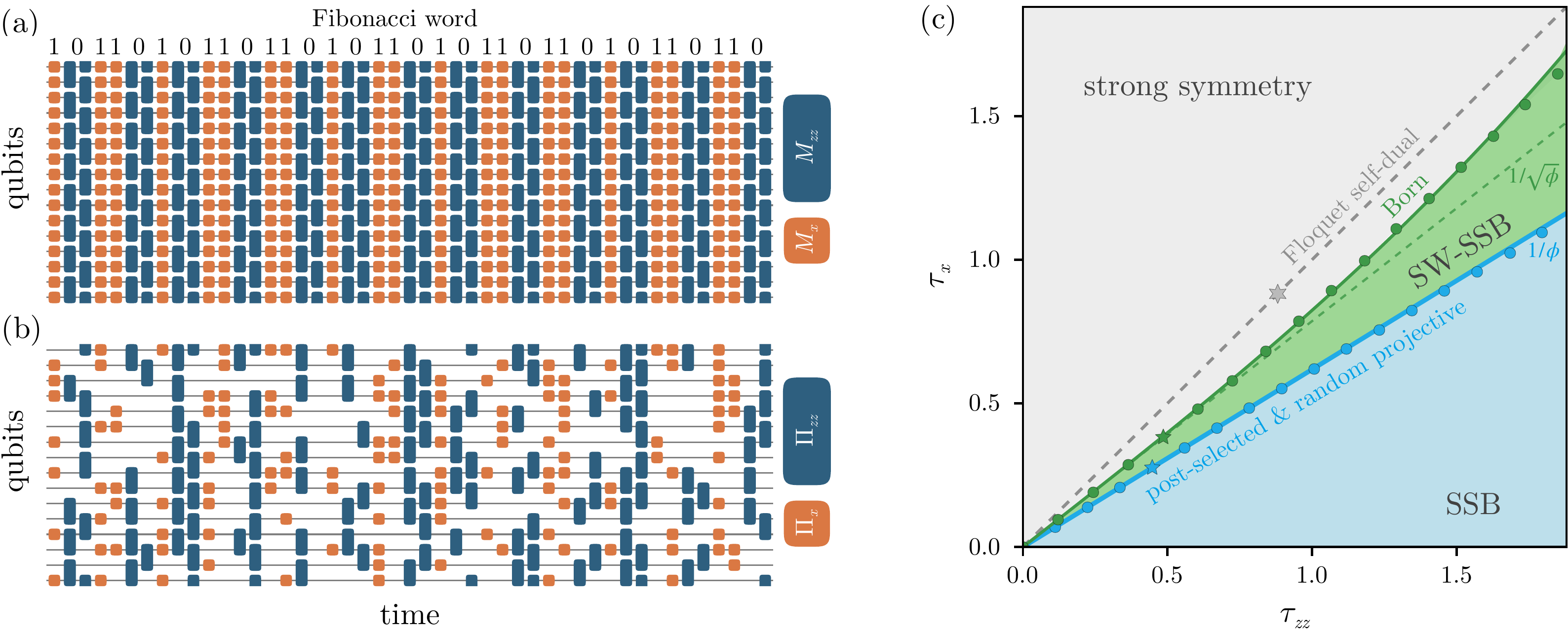}
    \caption{{\bf Quasiperiodically measured quantum Ising model.} 
    	(a) Fibonacci circuit of weak measurements using Kraus operators $M_x$ and $M_{zz}$ for $X$- and $ZZ$ measurements, respectively. 
		Measurement outcomes are post-selected in protocol (I) and Born-averaged in protocol (II). 
	(b) Random projective measurements with i.i.d.\ rates $p_x,p_{zz}$ (per layer) in the Clifford protocol (III). 
        (c) Phase diagrams as a function of measurement strengths $\tau_x$ and $\tau_{zz}$, extracted from the entanglement entropy. 
    	In the blue area, the 1D chain is long-range entangled for all three protocols. 
	The post-selected protocol gives rise to spontaneous symmetry-breaking (SSB), 
	while the Born and projective protocols exhibit weak breaking of the $\mathbb{Z}_2$ symmetry (strong-to-weak symmetry breaking, SW-SSB).
	In the gray area, the system is short-range entangled with unbroken strong $\mathbb{Z}_2$ symmetry.
	In the green area, the Born average trajectory is long-range entangled on average, 
	while the uniformly post-selected trajectory $s=-1$ remains short-range entangled, signaling SW-SSB.
        Blue data points are numerical results for the phase transition in the post-selected case (I),
        green data points numerical results for the Born-average model (II). 
        The blue and green lines indicate analytical estimates for the asymptotic form in the weak measurement limit (lower left corner).
        The diagonal gray dashed line shows the conventional Floquet self-dual location where the evolution times of $ZZ$ and $X$ are equal. 
         }
    \label{fig:QuantumCircuit}
\end{figure*}
%%%%%%%%%%%%%%%%%%%%%%%%%%%%%%%%%%%%%%%%%%%%%%%%%%%%%%%%%%%%%%%%%%%

In this manuscript, we characterize the steady-state phase transition and critical dynamics in prototypical {\it quasiperiodic} monitored Ising circuits without time-translation symmetry. The two competing measurements are arranged according to a ``Fibonacci word" as illustrated in Figs.~\ref{fig:QuantumCircuit}(a,b)~\cite{Dumitrescu2018,Maity2019,Lapierre2020,Wen_2021,Bhattacharjee2022,schmid24fibonacci,Lapierre2025,Kohmoto1983qpising,Siggia83qpising,Chandran17qpising,Crowley18qpising,PengRefael18,ElseHo20quasiperiodic},
interpolating between Floquet-periodic and fully random circuits. 
Apart from the quasi-random sequence of measurements, the dynamics exhibit inherent quantum mechanical randomness in the measurement outcomes, which determine the quantum trajectories of the state evolution.
We consider three distinct measurement protocols: (I) post-selection, (II) Born-averaged and (III) Clifford (projective) dynamics. 

The respective phase diagrams [Fig.~\ref{fig:QuantumCircuit}(c)] differ qualitatively in both the extent of the phases and the criticality.
With post-selection, the system dynamics are equivalently described as imaginary-time evolution with Ising-symmetric ``local Hamiltonians." The phase transition occurs when the ratio of the measurement strengths of the Ising interaction and the transverse field equals the golden ratio $\phi= (1+\sqrt{5})/2 = 1.618...$. This contrasts with Floquet circuits~\cite{Wang25selfdual}, where the transition occurs at a ratio of one. At the Fibonacci times, the effective circuit Hamiltonian is self-dual symmetric and described by the unitary Ising CFT in the long wavelength limit. For non-Fibonacci times, the quantum state exhibits measurement-altered Ising criticality \cite{Garratt22,Garratt23measureising,Alicea23measureising,jian23measureising,Sun2023,Paviglianiti2024}. For Born-averaged dynamics without post-selection, we find a parametrically enlarged long-range entangled (LRE) phase. The phase boundary is non-linear in the plane of measurement strengths, occurring at a ratio of $\sqrt{\phi}$ in the weak-measurement limit and near the Floquet ratio of one in the projective limit. Increasing measurement strength thus expands the LRE phase. In contrast, for the Clifford protocol, the phase boundary is essentially linear in the plane of measurement strengths and, surprisingly, agrees well with the phase transition line for post-selected trajectories.

%%%%%%%%%%%%%%%%%%%%%%%%%%%%%%%%%%%%%%%%%%%%%%%%%%%%%%%%%%%%%%%%%%%
% Models and protocols
%%%%%%%%%%%%%%%%%%%%%%%%%%%%%%%%%%%%%%%%%%%%%%%%%%%%%%%%%%%%%%%%%%%
\textit{Models and protocols.---} 
To create a quasiperiodic sequence of measurement operations, we consider binary Fibonacci words $w_n$, which concatenate the two previous words, $w_{n-1}w_{n-2}$. With the initial conditions $w_0=1$ and $w_1=10$, one generates the infinite sequence $w_\infty = 10110101...$ illustrated at the top of Fig.~\ref{fig:QuantumCircuit}(a). 
We then assemble the two competing measurements in our quantum circuits according to the sequence of binary values in  $w_\infty$,
translating 0 to $ZZ$ parity checks and 1 to $X$ measurements as shown in Figs.~\ref{fig:QuantumCircuit}(a) and (b).
The Kraus operators for local measurements at site $i$ are
\begin{align}
    M_x\propto e^{\tau_x  s_{x,i} X_i},
    \qquad 
    M_{zz}\propto e^{\tau_{zz}  s_{zz,i} Z_iZ_{i+1}}\ ,
    \label{eq:Metau}
\end{align}
where $\tau_{x(zz)}\in[0,\infty)$ quantifies the measurement strength of the $X_i$ and $Z_iZ_{i+1}$ measurements, respectively.
The binary measurement outcomes $s_{x,i}=\pm 1$ and $s_{zz,i}=\pm 1$ are Ising variables whose distributions follow Born's rule. Normalization factors $2\cosh(\tau_{x(zz)})$ are omitted in Eq.~\eqref{eq:Metau} for brevity.
The monitored Ising chain evolves according to the resulting product of Kraus operators,
\begin{align}
\ket{\Psi(\mathbf{s},t)}=M(\mathbf{s},t)\ket{\Psi(0)} \ ,
\end{align}
where $\mathbf{s}$ denotes the trajectory defined as the collection of measurement outcomes at all spatial locations and times, up to layer $t$. As the state is not normalized, the Born probability of a trajectory is given by the norm
\begin{equation}
    P(\mathbf{s}) = \braket{\Psi(\mathbf{s},t)}{\Psi(\mathbf{s},t)} \, .
\end{equation}
We initialize our circuit in the $+x$-polarized state  $\ket{\Psi(0)}=\ket{++\dots}$ with $X\ket{\pm}=+\ket{\pm}$,
which preserves the $\mathbb{Z}_2$ symmetry $\mathcal{P}= \prod_{j=1}^{L} X_j$
\footnote{
Such an initialization allows for the formation of LRE states in the SSB phase. 
More concretely, $\mathbb{Z}_2$ symmetry is preserved through the circuit, and thus the late-time state in the SSB phase 
corresponds to a Greenberger-Horne-Zeilinger type cat state, with long-range entanglement.
}.

Protocol (I) considers weak-measurement dynamics with uniform post-selection onto $s=+1$ for all measurements. In this conceptually simple case, the dynamics are equivalent to the discrete imaginary-time evolution of a quantum Ising chain under the Fibonacci drive.
Protocol (II) allows for random measurement outcomes according to Born's rule, so that specific measurement  trajectories are sampled with probability $P(\mathbf{s})$. 
Finally, protocol (III) [Fig.~\ref{fig:QuantumCircuit}(b)] considers the extreme limit of projective measurements ($\tau_{x(zz)}=\infty$), 
so that the circuit falls into the Clifford class~\cite{GottesmanKnill98,AaronsonGottesman04}.
In this protocol, we tune the dynamics  by randomly diluting the measurement gates. The projection operators $\Pi_x = (1\pm X_j)/2$ and $\Pi_{zz}= (1\pm Z_j Z_{j+1})/2$ are measured with probabilities $p_x$ and $p_{zz}$,
which are drawn independently and uniformly, see Fig.~\ref{fig:QuantumCircuit}(b). 
For a unifying phase diagram, we parametrize these measurement probabilities as
$p_{x,zz}=1-e^{-\tau_{x,zz}}$ in terms of an effective measurement strength $\tau$, which can be compared with the measurement strength in  protocols (I+II). 

%%%%%%%%%%%%%%%%%%%%%%%%%%%%%%%%%%%%%%%%%%%%%%%%%%%%%%%%%%%%%%%%%%%
\textit{ Phase diagrams.---} 
%
%%%%%%%%%%%%%%%%%%%%%%%%%%%%%%%%%%%%%%%%%%%%%%%%%%%%%%%%%%%%%%%%%%%
A joint {\it entanglement phase diagram} for all three protocols is shown in Fig.~\ref{fig:QuantumCircuit}(c). 
It has been mapped out by calculating the (Born-averaged) von Neumann entanglement entropy for a bipartition of the quantum chain
into two halves~\footnote{For a more precise determination of the critical locations we also adopt the ``coherent information'' as a diagnostic, which shows the best finite-size scaling data collapse performance~\cite{Wang25selfdual,Wan25nishimori, Eckstein25learning}}.
At late {\it Fibonacci times} this entropy is found to approach a steady value $S_{L/2}(\infty_F)$, saturating to a constant in the ``area-law'' phases, while diverging at the phase transition lines, exhibiting logarithmic scaling with the total system size. Between Fibonacci times, the entropy exhibits persistent fluctuations, which will be discussed below. Beyond the entanglement scaling, the phases are characterized by their behavior under the strong $\mathbb{Z}_2$ symmetry $\mathcal{P} \rho = \rho = \rho \mathcal{P}$ of the ensemble of post-measurement states with density matrix $\rho = \sum_{\mathbf{s}} P(\mathbf{s}) \ketbra{\Psi(\mathbf{s},t)}\otimes \ketbra{\mathbf{s}}$. The register state $\ket{\mathbf{s}}$ records the classical measurement outcomes, so that the trajectories can be unraveled. 
States in the regime $\tau_x\gg \tau_{zz}$ remain close to a product state $\ket{+}^{\otimes L}$ with unbroken strong Ising symmetry and $S_{L/2}(\infty)\simeq 0$.
States in the opposite regime $\tau_{zz} \gg \tau_{x}$ 
exhibit spontaneous symmetry breaking phenomena.
With post-selection (I), the system spontaneously breaks the strong symmetry, developing ferromagnetic order.  
In particular, at $\tau_x\simeq 0$ the dynamics  reaches a steady state $\frac{1}{\sqrt{2}}(\ket{\uparrow\uparrow\uparrow\dots}\pm \ket{\downarrow\downarrow\downarrow \dots})$, a ferromagnetic cat state~\footnote{Note that while the cat state maintains $\mathbb{Z}_2$ symmetry microscopically, in terms of phase of matter we still refer to it as $\mathbb{Z}_2$ SSB phase, defined by the long-range correlation}
 with non-vanishing connected quantum correlation $\langle Z_j Z_{j+\infty}\rangle_c = 1$. 
For the Born-average cases (II) and (III), the system only possesses spin-glass order due to the random measurement outcomes. 
For $\tau_x\simeq 0$, the ensemble of post-measurement states includes all random cat states with equal probabilities, e.g. $\frac{1}{\sqrt{2}}(\ket{\uparrow\uparrow\downarrow\dots}\pm \ket{\downarrow\downarrow\uparrow \dots})$. 
The long-range Ising correlations now have random signs, $\langle Z_j Z_{i}\rangle = \pm 1$, depending on the measurement outcomes. The measurement-averaged mixed state has {\it statistical average} long-range order 
\begin{equation}
[|\langle Z_j Z_{j+\infty}\rangle| ] \neq 0  \ ,\quad 
[\langle Z_j Z_{j+\infty}\rangle ] = 0 \ ,
\end{equation}
where $\langle \cdots \rangle $ denotes the quantum average of the spin, and $[\cdots]\equiv \sum_\mathbf{s} P(\mathbf{s})(\cdots)$ the statistical average over trajectories,  i.e.\ the classical measurement outcomes $\mathbf{s}$. 
This phase exhibits SW-SSB of the $\mathbb{Z}_2$ symmetry~\cite{Prosen12weaksym, Jiang14weaksym, Gorshkov20weaksym, Wang23averspt, Lee23decoher, Wang23aversym, Xu24higherformweaksym, You24weaksym, Luo24weaksym}, 
where the fidelity correlator~\cite{Wang24strtowksym} can be reduced~\cite{Wang25selfdual} to the spin-glass order parameter above: $\text{tr}\sqrt{\sqrt{\rho} Z_i Z_j \rho Z_i Z_j \sqrt{\rho}}
=
\sum_{\mathbf{s}} P(\mathbf{s}) |\langle Z_i Z_j\rangle| \neq 0$. 

%%%%%%%%%%%%%%%%%%%%%%%%%%%%%%%%%%%%%%%%%%%%%%%%%%%%%%%%%%%%%%%%%%%
\begin{figure}[b]
    \centering
    \includegraphics[width=.98\columnwidth]{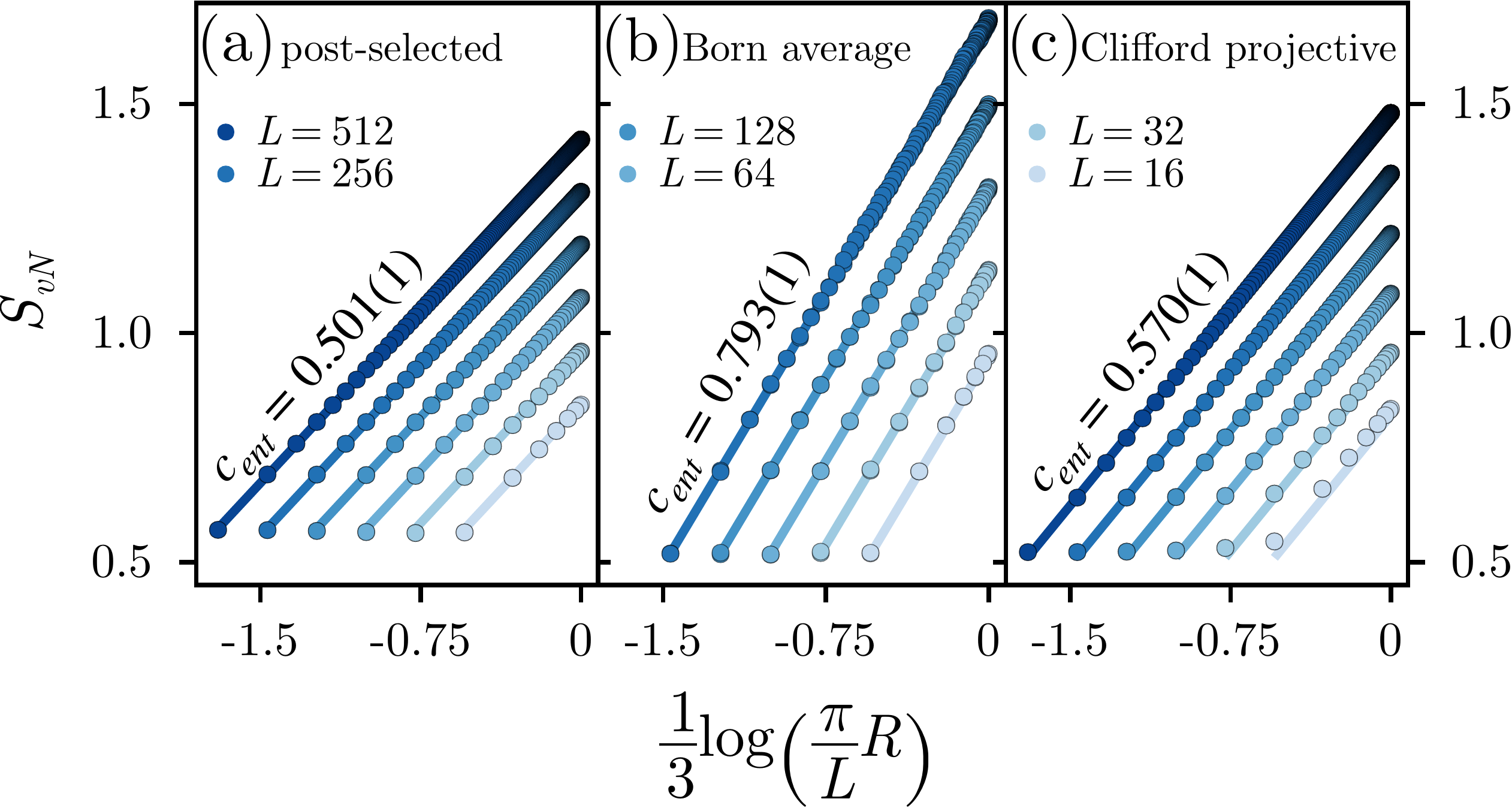}
    \caption{{\bf Monitored quantum criticality at late Fibonacci times.}
    	Shown is the entanglement entropy scaling for (a) post-selected trajectories, (b) Born trajectories, and (c) Clifford projective measurements
	as a function of the logarithmic chord distance $R(l,L) = \sin(\pi l/L)$. 
	The fitted slopes indicate the effective entanglement scaling dimension $c_{\rm ent}$ of the underlying transition.
	In the post-selected case (a) the estimated value 
	   $c_{\rm ent}=0.501(1)$ agrees with the Ising conformal field theory with central charge $1/2$. 
	   For the weak Born measurements (b) the estimated value $c_{\rm ent}=0.793(1)$ is in good agreement 
	   with the value $c_{\rm ent}=0.795(1)$ found for the Floquet circuit in Ref.~\cite{Wang25selfdual}.
	  For the Clifford projective measurements, our estimated value $c_{\rm ent}=0.570(1)$ matches the expectation for a percolation transition with 
	$c_{\rm ent} = 3\sqrt{3}/(2\pi)\ln2 = 0.573...$~\cite{Puetz24}. 
    }
    \label{fig:EntropyScaling_combined}
\end{figure}
%%%%%%%%%%%%%%%%%%%%%%%%%%%%%%%%%%%%%%%%%%%%%%%%%%%%%%%%%%%%%%%%%%%

The critical lines in the phase diagram differ between the three quasiperiodic protocols [Fig.~\ref{fig:QuantumCircuit}(c)], in contrast to the Floquet case where all three coincide at  $\tau_x=\tau_{zz}$. For post-selection (I) and the Clifford protocol (III), we find linear critical lines 
with a slope $\tau_x/\tau_{zz}=1/\phi$ given by the golden ratio. 
At smaller measurement ratios below this line 
[blue region in Fig.~\ref{fig:QuantumCircuit}(c)], the states are \textit{always} ordered, realizing SSB for the post-selection trajectories (I) and SW-SSB for the Born-averaged Clifford projective measurements (III). 
In contrast, for the Born weak measurement protocol (II), the critical line is non-linear. At small measurement strengths, its slope is approximately $\tau_{x}/\tau_{zz} \simeq 0.786 > 1/\phi$, while for large $\tau_{x,zz}$, it approaches the Floquet self-dual line $\tau_{x}/\tau_{zz} = 1$. The transition occurs at larger measurement ratios $\tau_x / \tau_{zz}$ than for (I) and (III), implying that weaker $ZZ$ Born measurements suffice to order the system. In the parameter regime between the two critical lines [green region in Fig.~\ref{fig:QuantumCircuit}(c) ], typical quantum states are ordered for Born weak measurements, but not for post-selected trajectories. 

We infer the universality class of the phase transitions by extracting the entanglement scaling dimension $c_{\rm ent}$~\footnote{Generally $c_{\rm ent}$ is the scaling dimension of the boundary condition changing operator~\cite{Ludwig2020}. For the special case of unitary conformal field theory (such as for the post-selected trajectory), $c_{\rm ent}$ is equivalent to the ``Casimir'' central charge determined via the free energy.} from fits of the Cardy-Calabrese ``entanglement arcs"~\cite{Calabrese2004,calabrese2009entanglement} 
(see End Matter). Our data in Fig.~\ref{fig:EntropyScaling_combined} shows three categories of $c_{\rm ent}$ consistent with: (I) an Ising transition with  $c_{\rm ent}=1/2$ for post-selection, (II) a non-unitary CFT transition with $c_{\rm ent}\approx 0.795$ for the Born weak measurement protocol~\cite{Wang25selfdual, Eckstein25learning}, and (III) bond percolation on the two-dimensional square lattice with $c_{\rm ent} \approx 0.573$~\cite{Buechler20, Puetz24} for the Clifford protocol. For all protocols, we find $c_{\rm ent}$ to remain unchanged along the entire transition line. 
Remarkably, these values all agree with those of the universality classes for the corresponding monitored Floquet circuits. While 
self-duality is an explicit symmetry in the Floquet case, quasiperiodic monitored Fibonacci circuits break time-translation symmetry and  self-duality is no longer present at the microscopic level. Our results indicate that self-duality is still an emergent property of the long-time evolution of the quasiperiodic circuit. This notion of ``emergence'' is similar in spirit to equilibrium settings where imaginary-time evolution leads to ground states which exhibit a larger set of symmetries than the Hamiltonian.\\

%%%%%%%%%%%%%%%%%%%%%%%%%%%%%%%%%%%%%%%%%%%%%%%%%%%%%%%%%%%%%%%%%%%
\textit{Analytical results.---} 
%%%%%%%%%%%%%%%%%%%%%%%%%%%%%%%%%%%%%%%%%%%%%%%%%%%%%%%%%%%%%%%%%%%
We provide a deeper understanding of the transitions and  universality classes via self-duality arguments. 
In protocol (I), recasting the circuit as a transfer matrix gives the Fibonacci recursion relation
\begin{align}
    M(f_{k-1})M(f_{k})=M(f_{k+1}) \, .
    \label{eq:Fib recursion}
\end{align}
This reflects the Fibonacci structure on the level of a single, clean trajectory. At small measurement strengths, the Kraus operator of the circuit can be written as $ M(t)\simeq e^{-tH_{\mathrm{eff}}}$, where we neglected all higher-order commutators in the high-frequency expansion. The effective Hamiltonian is a renormalized quantum Ising Hamiltonian, at Fibonacci times,
\begin{align}
    H_{\mathrm{eff}}\simeq \frac{\tau_x}{\phi}\sum_i X_i+\frac{\tau_{zz}}{\phi^2}\sum_i  Z_iZ_{i+1} \,.
\end{align}
The rescaled measurement strengths reflect the multiplicities of 
$X$ or $ZZ$ measurement gates in the circuit~\footnote{Notice that with our convention of the iterative rule $0\to 1,\ 1\to 10$, the ratio between the total number of ``1''s and ``0''s in a finite subsequence of the infinite Fibonacci word $10110101\cdots$ is asymptotically given by $\phi=1.618...$}. 
A duality transformation $\mathrm{KW}: X_j \to Z_{j-\frac{1}{2}}Z_{j+\frac{1}{2}},\ Z_j Z_{j+1}\to X_{j+\frac{1}{2}}$ exchanges the measurement layers, $M_x\leftrightarrow M_{zz}$. This maps the Fibonacci word to its dual word $\tilde{w}_\infty=1-w_\infty$, implying that the circuit is again Fibonacci with $1/\phi \to 1/\phi^2$. Correspondingly, the measurement strengths are mapped upon each other as
$\tau_x \leftrightarrow \tau_{zz}/\phi$. 
Therefore at $\tau_x=\tau_{zz}/\phi$ the model possesses~\cite{Wang25selfdual} a {\it strong}  self-duality symmetry 
\[
	\mathrm{KW}\rho=\rho \,.
\] 
This symmetry is {\it dynamical}, as it acts on the output state $\rho$ of the transfer matrix $M(f_k)$ at asymptotic Fibonacci times ($k \in \mathbb{N}$) rather than at the level of individual layers. The corresponding critical point is then described by the Ising CFT with central charge $c_{\rm ent}= 1/2$, 
consistent with our numerical finding, Fig.~\ref{fig:EntropyScaling_combined}(a). For a free-fermion perspective on this Ising transition, see End Matter.

In protocol (II), random measurement outcomes break translation invariance in space as well as the temporal Fibonacci structure of the Kraus operator in Eq.\ \eqref{eq:Fib recursion}. In the weak-measurement limit $\tau_{x,zz}\ll 1$, the binary measurement outcome has vanishing mean and unit variance, analogous to continuous weak measurements~\cite{Wang25selfdual} with measurement outcomes approximately following a Brownian motion~\cite{Schomerus2020weakmeasure, Ashida2020, Graham23majorana, Wang25selfdual}.  
We can then construct a Magnus expansion for a single realization of the circuit, $M(t)\simeq e^{-t H_{\mathrm{eff}}}$, with an effective Hamiltonian 
\begin{align}
    H_{\mathrm{eff}}
    \simeq \frac{\tau_x}{\sqrt{\phi}}\sum_{j} S_{x,j} X_j+
     \frac{\tau_{zz}}{\phi}\sum_{j}S_{zz,j}Z_jZ_{j+1} \, .
\label{eq:stat Ham}
\end{align}
This averages the sums over measurement outcomes over time,
$S_{x,j}=\frac{\sqrt{\phi}}{t}\sum^{t/\phi}_{n=1}s_{x,j}(n)$ and  $S_{zz,j}=\frac{\phi}{t}\sum^{t}_{n=t/\phi}s_{zz,j}(n)$. Both $S_{x,j}$ and $S_{zz,j}$ are normally distributed with zero mean and $1/\sqrt{t}$ width, reflecting self-averaging over the stochastic measurement record.  
This allows us to formulate a {\it statistical self-duality} \cite{Fisher1992,GRL2001,Wang25selfdual}, which applies to the ensemble of measurement trajectories rather than individual trajectories. A duality transformation exchanges the random variables $S_{x,j} \leftrightarrow S_{zz,j}$. The stochastic time-coarsening thus induces a {\it dynamical statistical self-duality}, provided that the variances of the rescaled random couplings match. This happens at criticality.  At the level of the post-measurement ensemble $\rho$, duality translates into a {\it weak symmetry} at asymptotic Fibonacci times \cite{Wang25selfdual} 
\[
	\mathrm{KW} \, \rho \, \mathrm{KW} = \rho \,.
\] 
At the same time,  $\mathrm{KW} \rho \neq \rho$ as some trajectories violate self-duality. For $\tau_{x,zz}\ll 1$, the resulting Hamiltonian is self-dual for $\tau_x /\tau_{zz}= 1/\sqrt{\phi} \simeq 0.786$. This estimate [dashed green line in Fig.\ \ref{fig:QuantumCircuit}(c)] agrees with our numerical results for weak measurement strengths. The next term in the Magnus expansion also yields an effective quantum Ising chain with renormalized couplings [see End Matter]. Statistical self-duality requires $\tau_x\sqrt{\phi} \simeq \tau_{zz} + \tau_{zz}^3/(8\phi^{2})$, which reproduces the boundary of the SW-SSB phase for larger $\tau_{zz}$ quite well [solid green line, Fig.\ \ref{fig:QuantumCircuit}(c)].
In the strong-measurement limit of  protocol (II), most qubits are effectively measured projectively. As repeated projective measurements of the same type do not change the state, $\left(\frac{1\pm X_j}{2}\right)^2 = \frac{1\pm X_j}{2}$, the circuit asymptotically approaches the corresponding Floquet case. As a result, the phase transition line of the Born weak measurement protocol approaches the Floquet self-dual transition line in the strong-measurement limit [Fig.\ \ref{fig:QuantumCircuit}(c)].  

Finally, in protocol (III), the dynamics maps onto anisotropic bond percolation on a square space--time lattice \cite{Buechler20} as illustrated in the End Matter. 

%%%%%%%%%%%%%%%%%%%%%%%%%%%%%%%%%%%%%%%%%%%%%%%%%%%%%%%%%%%%%%%%%%%
\begin{figure*}[tb]
    \centering
    \includegraphics[width=\textwidth]{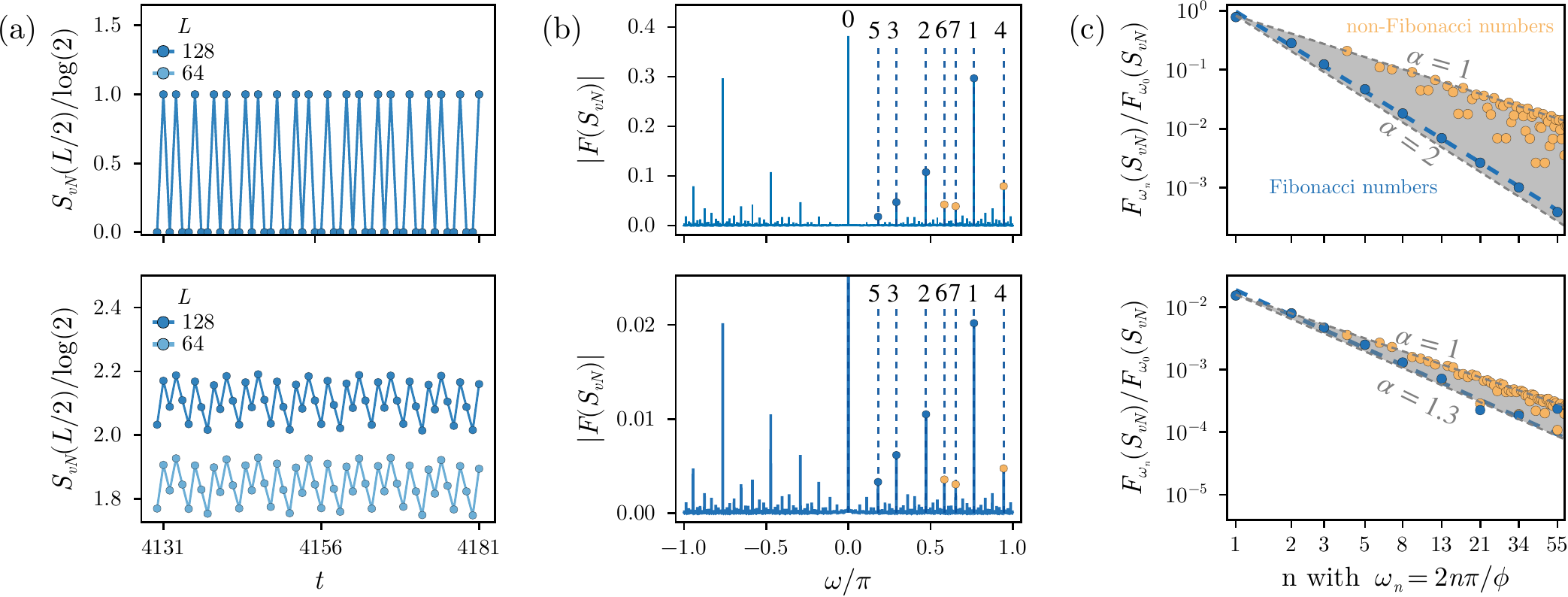}
    \caption{{\bf Entanglement dynamics and Fourier amplitudes in the Born-averaged protocol (II).} 
    	Top panels show the projective measurement limit where the state oscillates between a GHZ state with $\ln 2$ entropy 
	and a paramagnetic state with zero entropy, following the Fibonacci word; 
	the bottom panels are for finite measurement strength on the critical line  with $\tau_x\approx0.3823, \tau_{zz}\approx 0.4856$
    	(a) Half-cut entanglement entropy dynamics up to Fibonacci time $f_{19}=4181$. 
        (b) Fourier amplitudes of $S_{vN}(t)$, with frequency compactified within $[-\pi,\pi]$. 
        Dominant peaks  correspond to integer multiples of $\omega_n = 2n\pi/\phi$, such that the period is a fractional number of the golden ratio,
        $2\pi/\omega_n = \phi/n$. 
        The transform is performed for data within the time window $[f_{15}=610,f_{19}=4181]$. 
        (c) Scaling of the relative peak heights of Fourier amplitudes. 
        The amplitudes are lower- and upper-bounded by a power law decay $n^{-\alpha}$ with increasing $n$.}
    \label{fig:EntanglementDynamics}
\end{figure*}
%%%%%%%%%%%%%%%%%%%%%%%%%%%%%%%%%%%%%%%%%%%%%%%%%%%%%%%%%%%%%%%%%%%

%%%%%%%%%%%%%%%%%%%%%%%%%%%%%%%%%%%%%%%%%%%%%%%%%%%%%%%%%%%%%%%%%%%
\textit{Entanglement dynamics.---} 
%%%%%%%%%%%%%%%%%%%%%%%%%%%%%%%%%%%%%%%%%%%%%%%%%%%%%%%%%%%%%%%%%%%
Away from Fibonacci times, the quasiperiodicity of the circuit induces a {\it dynamically-deformed} criticality.
We analyze the entanglement evolution in the Born weak measurement protocol (II) at criticality in the ``steady" state. While the entropy saturates to a constant at Fibonacci times, it fluctuates at generic integer times. We first consider the projective-measurement limit. The time-evolved state is $\ket{+}^{\otimes L}$ after layers of completely packed $X$ measurements, and a random GHZ state after layers of completely packed $ZZ$ measurements. Thus its bipartite entanglement entropy $S_{vN}$ exactly follows the Fibonacci word [Fig.\ \ref{fig:EntanglementDynamics}(a), top].  
Its Fourier spectrum, $F[S_{vN}] (\omega)= \sum_{n} F_{\omega_n} [S_{vN}] \delta(\omega-\omega_n)$, has peaks at integer multiples of the golden ratio, $\omega_n=2\pi n/\phi$, [Fig.~\ref{fig:EntanglementDynamics}(b), top]. 
The amplitudes decay algebraically as $|F_{\omega_n} [S_{vN}]| \propto   \sin(n \pi/\phi^2)/n$, see App.\ \ref{sec:FT drive} and \cite{Levine1986}. Their scaling differs between Fibonacci ($n=f_k$) and non-Fibonacci harmonics. For non-Fibonacci harmonics, we find a $1/n$-power-law which reflects the square-pulse nature of the drive. In contrast, Fibonacci harmonics show a stronger $1/n^2$ scaling. This stronger decay originates from the oscillatory sine factor, which contributes an additional $1/n$-factor [Fig.~\ref{fig:EntanglementDynamics}(c), top].

At finite measurement strengths, the bipartite entanglement entropy $S_{vN}(t)$ shows weak quasiperiodic oscillations on top of the critical steady-state entanglement [Fig.~\ref{fig:EntanglementDynamics}(a), bottom]. The Fourier spectrum remains dominated by integer multiples of the golden ratio [Fig.~\ref{fig:EntanglementDynamics}(b), top], but generic Fibonacci and non-Fibonacci harmonics  obey the same power law $ |F_{\omega_n} [S_{vN}]|\propto   n^{-\alpha}$ [Fig.~\ref{fig:EntanglementDynamics}(c), bottom]. Decreasing the measurement strength continuously lowers $\alpha \leq 2$. 

The states beyond Fibonacci times can be understood as ``dynamical'' measurement-altered criticality with deformed $c_{\rm ent}$ which governs the logarithmically scaling entanglement entropy. As a consequence, all Fourier amplitudes are also expected to scale logarithmically with system size. As shown in End Matter, the zero frequency amplitude yields a scaling dimension $c_{\omega=0} \approx 0.790(1)$, close to the value for the Fibonacci times in Fig.~\ref{fig:EntropyScaling_combined}(b). \\

%%%%%%%%%%%%%%%%%%%%%%%%%%%%%%%%%%%%%%%%%%%%%%%%%%%%%%%%%%%%%%%%%%%
\textit{Outlook.---} 
%%%%%%%%%%%%%%%%%%%%%%%%%%%%%%%%%%%%%%%%%%%%%%%%%%%%%%%%%%%%%%%%%%%
We generalized monitored quantum circuits beyond the previously studied Floquet and fully random settings to time-quasiperiodic measurement protocols,
showcasing the Fibonacci sequence as a paradigmatic example. While unitarily driven systems generically heat to infinite temperature~\cite{DAlessio2014,Lazarides2014} (with notable exceptions such as prethermal or spectrally engineered drives~\cite{Abanin2017prethermal,Zhao2021multipolar,Zhao2023,Liu2026prethermalization}), measurements readily prevent the system from heating, enabling area-law scaling or critically entangled states. Quasiperiodicity enriches the measurement-induced dynamics and qualitatively modifies self-dual criticality. Critical monitored Floquet circuits are on average self-dual already after a single period. In contrast, in the quasiperiodic setting self-duality only emerges at the level of the full circuit sequence as a dynamical weak symmetry. 
Another generalization would extend the quasiperiodic structure to {\it space}, transforming the well-known static space-quasiperiodic Ising chain~\cite{Kohmoto1983qpising,Siggia83qpising,Chandran17qpising,Crowley18qpising} into a monitored version, in a spirit similar to lifting the Ising critical state to the weak self-dual critical state~\cite{Wang25selfdual}. 

Our results open a new direction for exploring the interplay between quasiperiodic driving and mid-circuit measurements, potentially enabling novel non-equilibrium phases of matter and new classes of dynamical quantum error correction codes. Beyond the influence of the quasiperiodic space-time  structure on many-body entanglement, it might be intriguing to study its impact on an encoded ``logical qubit'', or the classical measurement record,  when monitored quantum circuits are viewed as quantum error correction codes (with the Floquet code as an example~\cite{Haah21honeycomb,Hastings22honeycomb,Nat23floquetwithoutparent,Zhu23qubit}).
\\

{\it Data availability}.-- 
The numerical data shown in the figures is available on Zenodo~\cite{zenodo_fibonacci}.\\

%%%%%%%%%%%%%%%%%%%%%%%%%%%%%%%%%%%%%%%%%%%%%%%%%%%%%%%%%%%%%%%%%%%
% Acknowledgments
%%%%%%%%%%%%%%%%%%%%%%%%%%%%%%%%%%%%%%%%%%%%%%%%%%%%%%%%%%%%%%%%%%%
\begin{acknowledgments}
\textit{Acknowledgments.---}
We thank M. Knap and B. Lapierre for useful discussions.  
The Cologne research group is supported, in part, by
the Deutsche Forschungsgemeinschaft (DFG, German Research Foundation) under Germany’s Excellence Strategy—Cluster of Excellence Matter
and Light for Quantum Computing (ML4Q) EXC 2004/1 -- 390534769 
as well as within the CRC network TR 183 (Project Grant No.\ 277101999) as part of subproject B01.
Work at Freie Universit\"at Berlin was supported by Deutsche Forschungsgemeinschaft
through CRC 183 (project C03) and the German Excellence Strategy EXC3112/1 -- 533767171 (Center for Chiral Electronics). 
GYZ acknowledge the support of NSFC-Young Scientists Fund (grant no.~12504181) and Start-up Fund of HKUST(GZ) (grant no.~G0101000221), Guangdong provincial project (grant no.~2024QN11X201) and Guangdong Basic and Applied Basic Research Foundation (grant no.~2026A1515010965). 
Our numerical simulations were performed on the Noctua2 and Otus cluster at PC2 in Paderborn and the RAMSES cluster at RRZK Cologne. 
\end{acknowledgments}
%%%%%%%%%%%%%%%%%%%%%%%%%%%%%%%%%%%%%%%%%%%%%%%%%%%%%%%%%%%%%%%%%%%

%%%%%%%%%%%%%%%%%%%%%%%%%%%%%%%%%%%%%%%%%%%%%%%%%%%%%%%%%%%%%%%%%%%
\begin{center}
\bf End Matter
\end{center}
%%%%%%%%%%%%%%%%%%%%%%%%%%%%%%%%%%%%%%%%%%%%%%%%%%%%%%%%%%%%%%%%%%%

%%%%%%%%%%%%%%%%%%%%%%%%%%%%%%%%%%%%%%%%%%%%%%%%%%%%%%%%%%%%%%%%%%%
\subsection*{Extended data}
%%%%%%%%%%%%%%%%%%%%%%%%%%%%%%%%%%%%%%%%%%%%%%%%%%%%%%%%%%%%%%%%%%%

\begin{figure}[h!]
    \centering
      \includegraphics[width=0.98\columnwidth]{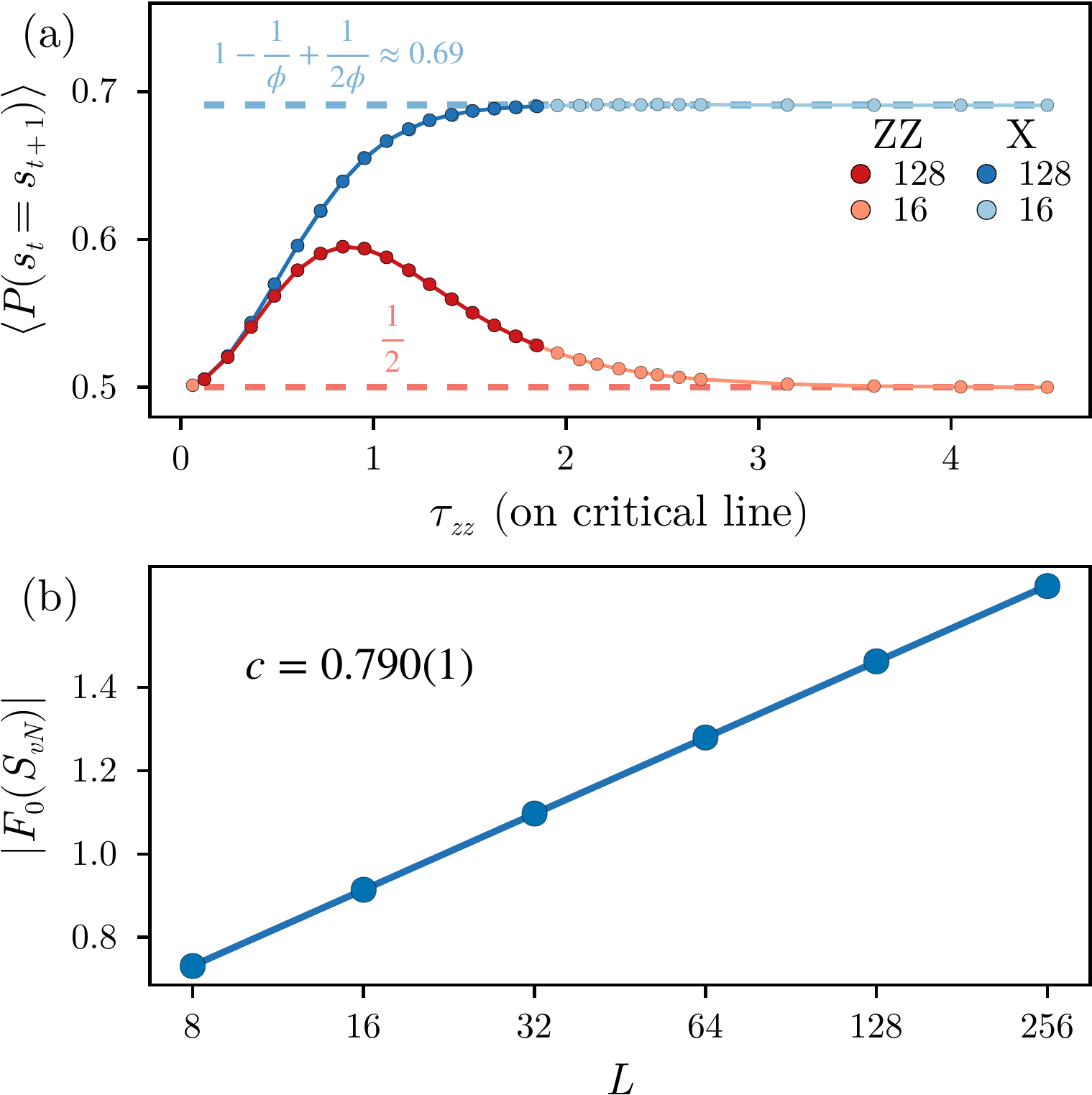}
    \caption{(a) {\bf Probability for a measurement correlation on the critical line.} 
    $P(s_t = s_{t+1})$ gives the probability that a measurement outcome stays the same in the next time step of the same kind. 
    $M_{zz}$ and $M_x$ have distinct behavior due to their different occurrence. Dashed lines mark the projective-limit expectations $P_{ZZ}=0.5$ and $P_{X}=1-1/\phi+1/(2\phi)$. 
    Data are for $L=128, 64, 32, 16$ on the critical line of the Born average; $L=64,32$ overlap with $L=128$.
    (b) {\bf Scaling of the Fourier peak at $\omega=0$ for weak measurement.}
   Parameters correspond to a point on the Born-averaged critical line with $\tau_x\!\approx\!0.3823, \tau_{zz}\!=\!0.4856$ (star in Fig.~\ref{fig:QuantumCircuit}(c)). 
   The entanglement scaling dimension $c_{\omega=0}=0.790(1)$ is close to $c_{\rm ent}=0.793(1)$ at Fibonacci times
   found in Fig.~\ref{fig:EntropyScaling_combined}. 
   Note that in Ref.~\cite{Wang25selfdual}, $c_{\rm ent}=0.795(1)$ for Floquet circuit of system sizes up to $L=512$. 
   }
    \label{fig:TimeCorrelationMeasurements}
\end{figure}

The measurement outcomes encode information about the quantum dynamics. Figure~\ref{fig:TimeCorrelationMeasurements}(a) shows the non-monotonic behavior of the nearest-neighbor temporal correlation of the measurement records as the measurement strength increases. Figure~\ref{fig:TimeCorrelationMeasurements}(b) shows the critical scaling of the zero-frequency Fourier amplitude of $S_{vN}(t)$. \\

%%%%%%%%%%%%%%%%%%%%%%%%%%%%%%%%%%%%%%%%%%%%%%%%%%%%%%%%%%%%%%%%%%%
%%%%%%%%%%%%%%%%%%%%%%%%%%%%%%%%%%%%%%%%%%%%%%%%%%%%%%%%%%%%%%%%%%%

%%%%%%%%%%%%%%%%%%%%%%%%%%%%%%%%%%%%%%%%%%%%%%%%%%%%%%%%%%%%%%%%%%%
\subsection*{Extended methods}
%%%%%%%%%%%%%%%%%%%%%%%%%%%%%%%%%%%%%%%%%%%%%%%%%%%%%%%%%%%%%%%%%%%

%%%%%%%%%%%%%%%%%%%%%%%%%%%%%%%%%%%%%%%%%%%%%%%%%%%%%%%%%%%%%%%%%%%
\subsubsection*{Entanglement arcs}
%%%%%%%%%%%%%%%%%%%%%%%%%%%%%%%%%%%%%%%%%%%%%%%%%%%%%%%%%%%%%%%%%%%

We fit the conformal field theory (CFT) prediction $S_{\mathrm{CFT}}(l) = \frac{c_{\rm ent}}{3}\ln\!\left[\frac{L}{\pi}\sin\!\left(\frac{\pi l}{L}\right)\right]$ \cite{Calabrese2004,calabrese2009entanglement} for periodic boundary conditions at criticality, where $c_{\rm ent}$ is the central charge for the unitary CFT but the scaling dimension of the boundary-condition-changing operator for measurement-induced criticality~\cite{Ludwig2020}. 

%%%%%%%%%%%%%%%%%%%%%%%%%%%%%%%%%%%%%%%%%%%%%%%%%%%%%%%%%%%%%%%%%%%
\subsubsection*{Protocol (I): free-fermion picture}
%%%%%%%%%%%%%%%%%%%%%%%%%%%%%%%%%%%%%%%%%%%%%%%%%%%%%%%%%%%%%%%%%%%

We analytically establish the transition in the post-selected protocol (I) via the gap-closing condition of the fermion representation. A Jordan-Wigner transformation maps the circuit to a Fibonacci-driven Kitaev chain in imaginary time \cite{schmid24fibonacci}. Translation symmetry labels modes by momentum $k$, and the single-particle Kraus operators $M_k(t)$ are $2\times 2$ matrices in particle-hole space (Pauli matrices $\sigma_i$). They follow the recursion Eq.~\eqref{eq:Fib recursion} with elementary transfer matrices
$\mathcal{M}_{zz}(k) \!\Leftrightarrow\! e^{i\frac{k}{2}\sigma^z}e^{2\tau_{zz} \sigma^y}e^{-i\frac{k}{2}\sigma^z}$, $\mathcal{M}_x(k) \!\Leftrightarrow\! e^{-2\tau_x \sigma^y}$. The transition follows from gap closings of the single-particle relaxation spectrum $\Gamma_k$ in $M_k(t)=e^{-\Gamma_k t \sigma^z }$, occurring at $k=0$. The spectrum is gapless for $\tau_x/\tau_{zz}<1/\phi$ and gapped otherwise, signaling a change in steady-state degeneracy and hence in entanglement. At criticality, the spectrum is linear around $k\simeq 0$, producing a Dirac cone $\Gamma_{k} \simeq \pm v k$ with $v = \sinh\!\left(2\tau_{x} / \phi \right)$, consistent with the post-selected dynamics remaining in the Ising universality class for all measurement strengths.

%%%%%%%%%%%%%%%%%%%%%%%%%%%%%%%%%%%%%%%%%%%%%%%%%%%%%%%%%%%%%%%%%%%
\begin{figure*}[t!]
    \centering
    \includegraphics[width=\textwidth]{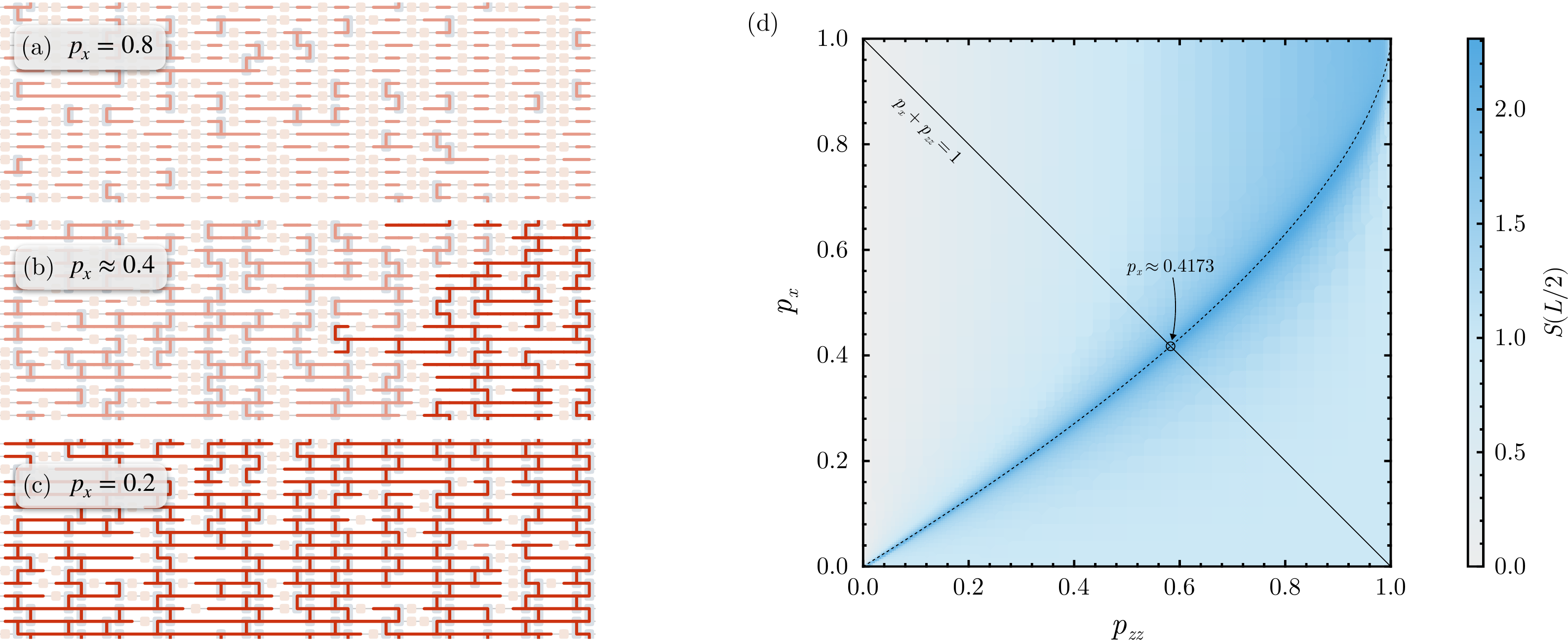}
    \caption{{\bf Bond-percolation transition in space-time (protocol III).}
    (a--c) A $ZZ$ projective measurement entangles two neighboring qubits and corresponds to a vertical link gluing the two sites; 
    the qubit world lines stretch left to right. An $X$ projective measurement disentangles a qubit and corresponds to a broken horizontal link.
    (d) Random projective-measurement phase diagram in $p$-space. 
    The solid line marks $p_x+p_{zz}=1$, while the dashed line shows $\tau_{x} = \tau_{zz}/\phi$ projected into $p$-space via $p=1-e^{-\tau}$, 
    visualizing the behavior at infinite $\tau$. Simulation depth is $\mathrm{fib}_{17}$, averaged over $98,304$ random trajectories.}
    \label{fig:Punchcard_percolation}
\end{figure*}
%%%%%%%%%%%%%%%%%%%%%%%%%%%%%%%%%%%%%%%%%%%%%%%%%%%%%%%%%%%%%%%%%%%

%%%%%%%%%%%%%%%%%%%%%%%%%%%%%%%%%%%%%%%%%%%%%%%%%%%%%%%%%%%%%%%%%%%
\subsubsection*{Protocol (II): Magnus expansion}
%%%%%%%%%%%%%%%%%%%%%%%%%%%%%%%%%%%%%%%%%%%%%%%%%%%%%%%%%%%%%%%%%%%

We derive a Magnus expansion for the effective Hamiltonian in the weak Born protocol (II). The transfer matrix of a single trajectory up to time $t$ is $M(\mathbf{s},t)= \prod^t_{n=1}M_{w_n}(\{s_j(n)\})$, where $w_n=0,1$ denotes an $X,ZZ$ measurement layer with outcomes $\{s_j(n)\}$. The Born probability for outcome $s$ of operator $O=X_i,Z_iZ_{i+1}$ at strength $\tau$ is $p(s) = \frac{1}{2}(1+s \tanh \tau\expval{O}(t))$. In the weak-measurement regime $\tau\ll 1$, the measurements are approximately unbiased and temporally uncorrelated (numerically corroborated in Figs.~\ref{fig:TimeCorrelationMeasurements}(a) and \ref{fig:Disorder_pureaverage}). 

The weak-measurement limit admits a Magnus expansion~\cite{Magnus1954,Blanes2009magnus} $M(t)=e^{-tH_{\mathrm{eff}}}$ with $H_{\mathrm{eff}}=\sum^\infty_{l=0}H_l$ and $H_l=\mathcal{O}(\tau^l)$, derived for quasiperiodic Fibonacci driving in Ref.~\cite{Dumitrescu2018}. Keeping the first two orders yields
\begin{align}
    H_0 &= \frac{\tau_x}{\sqrt{\phi}}\sum_{j}S_{x,j} X_j + \frac{\tau_{zz}}{\phi}\sum_{j} S_{zz,j}Z_j Z_{j+1},
    \\
    H_1 &= \frac{\tau_x \tau_{zz}}{2}\sum_{j} R_{xz,j}  [X_j,Z_j Z_{j+1}],
\end{align}
with the (bare) sums over measurement outcomes $S_{x,j}=\frac{\sqrt{\phi}}{t}\sum^t_n w_n s_{j}(n)$ and $S_{zz,j}=\frac{\phi}{t}\sum^t_n(1-w_n) s_{j}(n)$. The cross term $R_{xz,j}=\frac{1}{t}\sum_{n,m<n} s_j(n) s_j(m) (w_n-w_m)$ represents the first correction to the time average. The expansion is controlled for sufficiently short times satisfying $\ln \norm{M(\mathbf{s},t)} <1 $, which defines a stopping time $t^\ast=1/\max(\tau_{x,zz})$, sufficient to coarse-grain the dynamics. Evaluating the commutator and locally rotating around the $x$-axis diagonalizes $H_0+H_1$ into an effective transverse-field Ising model with rescaled couplings $\tilde{S}_{zz,j}\simeq S_{zz,j}+(\tau_x \phi R_{xz,j})^2/S_{x,j}$.

The phase transition follows from statistical self-duality. At leading order, the bare measurement sums satisfy $\langle S^2_{x,j} \rangle = \langle S^2_{zz,j} \rangle =1/t$, yielding $\tau_x /\tau_{zz}\simeq 1/\sqrt{\phi}$. Including the next-order correction via the cross-term width $\langle R_{xz,j}^2\rangle= 1/(8 \phi^3)$ and replacing random variables by their widths reproduces the critical line quoted in the main text.

%%%%%%%%%%%%%%%%%%%%%%%%%%%%%%%%%%%%%%%%%%%%%%%%%%%%%%%%%%%%%%%%%%%
\subsubsection*{Protocol (III): bond-percolation transition in space-time}
%%%%%%%%%%%%%%%%%%%%%%%%%%%%%%%%%%%%%%%%%%%%%%%%%%%%%%%%%%%%%%%%%%%

The Clifford protocol (III) maps to anisotropic bond percolation on a square space-time lattice [Fig.~\ref{fig:Punchcard_percolation}(a-c)]: $ZZ$ ($X$) measurements activate vertical (horizontal) bonds, and the entanglement of the final state tracks the Bell cluster spanning a percolating cluster in space-time. Spatial translation symmetry holds on average, while time-translation symmetry is broken by the Fibonacci structure.

The critical line can then be obtained using {\it statistical} duality arguments for bond percolation \cite{Sykes1963}. Taking into account the probability of the recurrence of $X$ measurements in the Fibonacci word, we find an analytic expression for the self-dual critical location
\begin{align}
    \frac{p_{zz}}{\phi}= p_x - \frac{p_x^2}{2\phi}+\order{p_x^3} \,.
    \label{eq:pert theory}
\end{align}
The leading correction accounts for repeated $X$ measurements where the bond is already activated, occurring with probability $\sim p_x^2$. The Fibonacci structure is key here: only the segment ``11'' appears in the Fibonacci word, never ``00'' [Fig.~\ref{fig:QuantumCircuit}(a)], so only repeated $X$ measurements are counted, giving Eq.~\eqref{eq:pert theory}. Higher-order corrections from longer correlated sequences are suppressed by additional powers of $1/\phi$. In the opposite limit $p_x,p_{zz}\to1$, the line terminates at $p_x=p_{zz}=1$. Figure~\ref{fig:Punchcard_percolation}(d) shows the phase diagram in $(p_x,p_{zz})$-space; the boundary $p_{zz}=1-(1-p_{x})^\phi$, inferred from the linear $\tau_x=\tau_{zz}/\phi$ relation, reproduces Eq.~\eqref{eq:pert theory} upon expansion for small $p_x$.

The critical behavior of protocol (III) lies in the universality class of two-dimensional bond percolation with central charge $c_{\rm ent}=\frac{3\sqrt{3}}{2\pi}\ln(2)$, which is consistent with our numerics. Despite the absence of time-translation symmetry, the long-wavelength behavior matches the Floquet case, indicating that quasiperiodicity is irrelevant in the renormalization-group sense at criticality.

%%%%%%%%%%%%%%%%%%%%%%%%%%%%%%%%%%%%%%%%%%%%%%%%%%%%%%%%%%%%%%%%%%%
% Bibliography
%%%%%%%%%%%%%%%%%%%%%%%%%%%%%%%%%%%%%%%%%%%%%%%%%%%%%%%%%%%%%%%%%%%
\bibliography{measurements}

\begin{thebibliography}{82}%
\makeatletter
\providecommand \@ifxundefined [1]{%
 \@ifx{#1\undefined}
}%
\providecommand \@ifnum [1]{%
 \ifnum #1\expandafter \@firstoftwo
 \else \expandafter \@secondoftwo
 \fi
}%
\providecommand \@ifx [1]{%
 \ifx #1\expandafter \@firstoftwo
 \else \expandafter \@secondoftwo
 \fi
}%
\providecommand \natexlab [1]{#1}%
\providecommand \enquote  [1]{``#1''}%
\providecommand \bibnamefont  [1]{#1}%
\providecommand \bibfnamefont [1]{#1}%
\providecommand \citenamefont [1]{#1}%
\providecommand \href@noop [0]{\@secondoftwo}%
\providecommand \href [0]{\begingroup \@sanitize@url \@href}%
\providecommand \@href[1]{\@@startlink{#1}\@@href}%
\providecommand \@@href[1]{\endgroup#1\@@endlink}%
\providecommand \@sanitize@url [0]{\catcode `\\12\catcode `\$12\catcode `\&12\catcode `\#12\catcode `\^12\catcode `\_12\catcode `\%12\relax}%
\providecommand \@@startlink[1]{}%
\providecommand \@@endlink[0]{}%
\providecommand \url  [0]{\begingroup\@sanitize@url \@url }%
\providecommand \@url [1]{\endgroup\@href {#1}{\urlprefix }}%
\providecommand \urlprefix  [0]{URL }%
\providecommand \Eprint [0]{\href }%
\providecommand \doibase [0]{https://doi.org/}%
\providecommand \selectlanguage [0]{\@gobble}%
\providecommand \bibinfo  [0]{\@secondoftwo}%
\providecommand \bibfield  [0]{\@secondoftwo}%
\providecommand \translation [1]{[#1]}%
\providecommand \BibitemOpen [0]{}%
\providecommand \bibitemStop [0]{}%
\providecommand \bibitemNoStop [0]{.\EOS\space}%
\providecommand \EOS [0]{\spacefactor3000\relax}%
\providecommand \BibitemShut  [1]{\csname bibitem#1\endcsname}%
\let\auto@bib@innerbib\@empty
%</preamble>
\bibitem [{\citenamefont {Fisher}\ \emph {et~al.}(2023)\citenamefont {Fisher}, \citenamefont {Khemani}, \citenamefont {Nahum},\ and\ \citenamefont {Vijay}}]{Fisher2022reviewMIPT}%
  \BibitemOpen
  \bibfield  {author} {\bibinfo {author} {\bibfnamefont {M.~P.}\ \bibnamefont {Fisher}}, \bibinfo {author} {\bibfnamefont {V.}~\bibnamefont {Khemani}}, \bibinfo {author} {\bibfnamefont {A.}~\bibnamefont {Nahum}},\ and\ \bibinfo {author} {\bibfnamefont {S.}~\bibnamefont {Vijay}},\ }\bibfield  {title} {\bibinfo {title} {{Random Quantum Circuits}},\ }\href {https://doi.org/https://doi.org/10.1146/annurev-conmatphys-031720-030658} {\bibfield  {journal} {\bibinfo  {journal} {Annu. Rev. Condens. Matter Phys.}\ }\textbf {\bibinfo {volume} {14}},\ \bibinfo {pages} {335} (\bibinfo {year} {2023})}\BibitemShut {NoStop}%
\bibitem [{\citenamefont {Potter}\ and\ \citenamefont {Vasseur}(2022)}]{Potter21review}%
  \BibitemOpen
  \bibfield  {author} {\bibinfo {author} {\bibfnamefont {A.~C.}\ \bibnamefont {Potter}}\ and\ \bibinfo {author} {\bibfnamefont {R.}~\bibnamefont {Vasseur}},\ }\bibinfo {title} {{Entanglement Dynamics in Hybrid Quantum Circuits}},\ in\ \href@noop {} { {\bibinfo {booktitle} {\it Entanglement in Spin Chains: From Theory to Quantum Technology Applications}}},\ \bibinfo {editor} {edited by\ \bibinfo {editor} {\bibfnamefont {A.}~\bibnamefont {Bayat}}, \bibinfo {editor} {\bibfnamefont {S.}~\bibnamefont {Bose}},\ and\ \bibinfo {editor} {\bibfnamefont {H.}~\bibnamefont {Johannesson}}}\ (\bibinfo  {publisher} {Springer International Publishing},\ \bibinfo {address} {Cham},\ \bibinfo {year} {2022})\ pp.\ \bibinfo {pages} {211--249}\BibitemShut {NoStop}%
\bibitem [{\citenamefont {Lang}\ and\ \citenamefont {B\"uchler}(2020)}]{Buechler20}%
  \BibitemOpen
  \bibfield  {author} {\bibinfo {author} {\bibfnamefont {N.}~\bibnamefont {Lang}}\ and\ \bibinfo {author} {\bibfnamefont {H.~P.}\ \bibnamefont {B\"uchler}},\ }\bibfield  {title} {\bibinfo {title} {{Entanglement transition in the projective transverse field Ising model}},\ }\href {https://doi.org/10.1103/PhysRevB.102.094204} {\bibfield  {journal} {\bibinfo  {journal} {Phys. Rev. B}\ }\textbf {\bibinfo {volume} {102}},\ \bibinfo {pages} {094204} (\bibinfo {year} {2020})}\BibitemShut {NoStop}%
\bibitem [{\citenamefont {Sang}\ and\ \citenamefont {Hsieh}(2021)}]{Hsieh2021measure}%
  \BibitemOpen
  \bibfield  {author} {\bibinfo {author} {\bibfnamefont {S.}~\bibnamefont {Sang}}\ and\ \bibinfo {author} {\bibfnamefont {T.~H.}\ \bibnamefont {Hsieh}},\ }\bibfield  {title} {\bibinfo {title} {{Measurement-protected quantum phases}},\ }\href {https://doi.org/10.1103/PhysRevResearch.3.023200} {\bibfield  {journal} {\bibinfo  {journal} {Phys. Rev. Res.}\ }\textbf {\bibinfo {volume} {3}},\ \bibinfo {pages} {023200} (\bibinfo {year} {2021})}\BibitemShut {NoStop}%
\bibitem [{\citenamefont {Pfeuty}(1970)}]{Pfeuty1970}%
  \BibitemOpen
  \bibfield  {author} {\bibinfo {author} {\bibfnamefont {P.}~\bibnamefont {Pfeuty}},\ }\bibfield  {title} {\bibinfo {title} {{The one-dimensional Ising model with a transverse field}},\ }\href {https://doi.org/https://doi.org/10.1016/0003-4916(70)90270-8} {\bibfield  {journal} {\bibinfo  {journal} {Ann. Phys.}\ }\textbf {\bibinfo {volume} {57}},\ \bibinfo {pages} {79} (\bibinfo {year} {1970})}\BibitemShut {NoStop}%
\bibitem [{\citenamefont {Sachdev}(2011)}]{Sachdev2011}%
  \BibitemOpen
  \bibfield  {author} {\bibinfo {author} {\bibfnamefont {S.}~\bibnamefont {Sachdev}},\ }\href@noop {} { {\bibinfo {title} {{\it Quantum phase transitions}}}},\ \bibinfo {edition} {2nd}\ ed.\ (\bibinfo  {publisher} {Cambridge University Press},\ \bibinfo {address} {Cambridge},\ \bibinfo {year} {2011})\BibitemShut {NoStop}%
\bibitem [{\citenamefont {{Wang}}\ \emph {et~al.}(2025)\citenamefont {{Wang}}, \citenamefont {{Vasseur}}, \citenamefont {{Trebst}}, \citenamefont {{Ludwig}},\ and\ \citenamefont {{Zhu}}}]{Wang25selfdual}%
  \BibitemOpen
  \bibfield  {author} {\bibinfo {author} {\bibfnamefont {Q.}~\bibnamefont {{Wang}}}, \bibinfo {author} {\bibfnamefont {R.}~\bibnamefont {{Vasseur}}}, \bibinfo {author} {\bibfnamefont {S.}~\bibnamefont {{Trebst}}}, \bibinfo {author} {\bibfnamefont {A.~W.~W.}\ \bibnamefont {{Ludwig}}},\ and\ \bibinfo {author} {\bibfnamefont {G.-Y.}\ \bibnamefont {{Zhu}}},\ }\bibfield  {title} {\bibinfo {title} {{Decoherence-induced self-dual criticality in topological states of matter}},\ }\href@noop {} {\bibfield  {journal} {\bibinfo  {journal} {preprint}\ } (\bibinfo {year} {2025})},\ \Eprint {https://arxiv.org/abs/2502.14034} {arXiv:2502.14034} \BibitemShut {NoStop}%
\bibitem [{\citenamefont {Lessa}\ \emph {et~al.}(2025)\citenamefont {Lessa}, \citenamefont {Ma}, \citenamefont {Zhang}, \citenamefont {Bi}, \citenamefont {Cheng},\ and\ \citenamefont {Wang}}]{Wang24strtowksym}%
  \BibitemOpen
  \bibfield  {author} {\bibinfo {author} {\bibfnamefont {L.~A.}\ \bibnamefont {Lessa}}, \bibinfo {author} {\bibfnamefont {R.}~\bibnamefont {Ma}}, \bibinfo {author} {\bibfnamefont {J.-H.}\ \bibnamefont {Zhang}}, \bibinfo {author} {\bibfnamefont {Z.}~\bibnamefont {Bi}}, \bibinfo {author} {\bibfnamefont {M.}~\bibnamefont {Cheng}},\ and\ \bibinfo {author} {\bibfnamefont {C.}~\bibnamefont {Wang}},\ }\bibfield  {title} {\bibinfo {title} {{Strong-to-Weak Spontaneous Symmetry Breaking in Mixed Quantum States}},\ }\href {https://doi.org/10.1103/PRXQuantum.6.010344} {\bibfield  {journal} {\bibinfo  {journal} {PRX Quantum}\ }\textbf {\bibinfo {volume} {6}},\ \bibinfo {pages} {010344} (\bibinfo {year} {2025})}\BibitemShut {NoStop}%
\bibitem [{\citenamefont {{Zhang}}\ \emph {et~al.}(2024)\citenamefont {{Zhang}}, \citenamefont {{Xu}}, \citenamefont {{Zhang}}, \citenamefont {{Xu}}, \citenamefont {{Bi}},\ and\ \citenamefont {{Luo}}}]{Luo24weaksym}%
  \BibitemOpen
  \bibfield  {author} {\bibinfo {author} {\bibfnamefont {C.}~\bibnamefont {{Zhang}}}, \bibinfo {author} {\bibfnamefont {Y.}~\bibnamefont {{Xu}}}, \bibinfo {author} {\bibfnamefont {J.-H.}\ \bibnamefont {{Zhang}}}, \bibinfo {author} {\bibfnamefont {C.}~\bibnamefont {{Xu}}}, \bibinfo {author} {\bibfnamefont {Z.}~\bibnamefont {{Bi}}},\ and\ \bibinfo {author} {\bibfnamefont {Z.-X.}\ \bibnamefont {{Luo}}},\ }\bibfield  {title} {\bibinfo {title} {{Strong-to-weak spontaneous breaking of 1-form symmetry and intrinsically mixed topological order}},\ }\href@noop {} {\bibfield  {journal} {\bibinfo  {journal} {preprint}\ } (\bibinfo {year} {2024})},\ \Eprint {https://arxiv.org/abs/2409.17530} {arXiv:2409.17530} \BibitemShut {NoStop}%
\bibitem [{\citenamefont {Sala}\ \emph {et~al.}(2024)\citenamefont {Sala}, \citenamefont {Gopalakrishnan}, \citenamefont {Oshikawa},\ and\ \citenamefont {You}}]{You24weaksym}%
  \BibitemOpen
  \bibfield  {author} {\bibinfo {author} {\bibfnamefont {P.}~\bibnamefont {Sala}}, \bibinfo {author} {\bibfnamefont {S.}~\bibnamefont {Gopalakrishnan}}, \bibinfo {author} {\bibfnamefont {M.}~\bibnamefont {Oshikawa}},\ and\ \bibinfo {author} {\bibfnamefont {Y.}~\bibnamefont {You}},\ }\bibfield  {title} {\bibinfo {title} {{Spontaneous strong symmetry breaking in open systems: Purification perspective}},\ }\href {https://doi.org/10.1103/PhysRevB.110.155150} {\bibfield  {journal} {\bibinfo  {journal} {Phys. Rev. B}\ }\textbf {\bibinfo {volume} {110}},\ \bibinfo {pages} {155150} (\bibinfo {year} {2024})}\BibitemShut {NoStop}%
\bibitem [{\citenamefont {Sun}\ \emph {et~al.}(2025)\citenamefont {Sun}, \citenamefont {Zhang},\ and\ \citenamefont {Feng}}]{Sun2025}%
  \BibitemOpen
  \bibfield  {author} {\bibinfo {author} {\bibfnamefont {N.}~\bibnamefont {Sun}}, \bibinfo {author} {\bibfnamefont {P.}~\bibnamefont {Zhang}},\ and\ \bibinfo {author} {\bibfnamefont {L.}~\bibnamefont {Feng}},\ }\bibfield  {title} {\bibinfo {title} {{Scheme to Detect the Strong-to-Weak Symmetry Breaking via Randomized Measurements}},\ }\href {https://doi.org/10.1103/7p5x-7yqb} {\bibfield  {journal} {\bibinfo  {journal} {Phys. Rev. Lett.}\ }\textbf {\bibinfo {volume} {135}},\ \bibinfo {pages} {090403} (\bibinfo {year} {2025})}\BibitemShut {NoStop}%
\bibitem [{\citenamefont {Ziereis}\ \emph {et~al.}(2025)\citenamefont {Ziereis}, \citenamefont {Moudgalya},\ and\ \citenamefont {Knap}}]{Ziereis2025}%
  \BibitemOpen
  \bibfield  {author} {\bibinfo {author} {\bibfnamefont {N.}~\bibnamefont {Ziereis}}, \bibinfo {author} {\bibfnamefont {S.}~\bibnamefont {Moudgalya}},\ and\ \bibinfo {author} {\bibfnamefont {M.}~\bibnamefont {Knap}},\ }\href {https://arxiv.org/abs/2509.09669} {\bibinfo {title} {{Strong-to-Weak Symmetry Breaking Phases in Steady States of Quantum Operations}}} (\bibinfo {year} {2025}),\ \Eprint {https://arxiv.org/abs/2509.09669} {arXiv:2509.09669 [cond-mat.stat-mech]} \BibitemShut {NoStop}%
\bibitem [{\citenamefont {Jian}\ \emph {et~al.}(2020)\citenamefont {Jian}, \citenamefont {You}, \citenamefont {Vasseur},\ and\ \citenamefont {Ludwig}}]{Ludwig2020}%
  \BibitemOpen
  \bibfield  {author} {\bibinfo {author} {\bibfnamefont {C.-M.}\ \bibnamefont {Jian}}, \bibinfo {author} {\bibfnamefont {Y.-Z.}\ \bibnamefont {You}}, \bibinfo {author} {\bibfnamefont {R.}~\bibnamefont {Vasseur}},\ and\ \bibinfo {author} {\bibfnamefont {A.~W.~W.}\ \bibnamefont {Ludwig}},\ }\bibfield  {title} {\bibinfo {title} {{Measurement-induced criticality in random quantum circuits}},\ }\href {https://doi.org/10.1103/PhysRevB.101.104302} {\bibfield  {journal} {\bibinfo  {journal} {Phys. Rev. B}\ }\textbf {\bibinfo {volume} {101}},\ \bibinfo {pages} {104302} (\bibinfo {year} {2020})}\BibitemShut {NoStop}%
\bibitem [{\citenamefont {Bao}\ \emph {et~al.}(2020)\citenamefont {Bao}, \citenamefont {Choi},\ and\ \citenamefont {Altman}}]{Bao20learning}%
  \BibitemOpen
  \bibfield  {author} {\bibinfo {author} {\bibfnamefont {Y.}~\bibnamefont {Bao}}, \bibinfo {author} {\bibfnamefont {S.}~\bibnamefont {Choi}},\ and\ \bibinfo {author} {\bibfnamefont {E.}~\bibnamefont {Altman}},\ }\bibfield  {title} {\bibinfo {title} {{Theory of the phase transition in random unitary circuits with measurements}},\ }\href {https://doi.org/10.1103/PhysRevB.101.104301} {\bibfield  {journal} {\bibinfo  {journal} {Phys. Rev. B}\ }\textbf {\bibinfo {volume} {101}},\ \bibinfo {pages} {104301} (\bibinfo {year} {2020})}\BibitemShut {NoStop}%
\bibitem [{\citenamefont {Zabalo}\ \emph {et~al.}(2020)\citenamefont {Zabalo}, \citenamefont {Gullans}, \citenamefont {Wilson}, \citenamefont {Gopalakrishnan}, \citenamefont {Huse},\ and\ \citenamefont {Pixley}}]{Pixley20mipt}%
  \BibitemOpen
  \bibfield  {author} {\bibinfo {author} {\bibfnamefont {A.}~\bibnamefont {Zabalo}}, \bibinfo {author} {\bibfnamefont {M.~J.}\ \bibnamefont {Gullans}}, \bibinfo {author} {\bibfnamefont {J.~H.}\ \bibnamefont {Wilson}}, \bibinfo {author} {\bibfnamefont {S.}~\bibnamefont {Gopalakrishnan}}, \bibinfo {author} {\bibfnamefont {D.~A.}\ \bibnamefont {Huse}},\ and\ \bibinfo {author} {\bibfnamefont {J.~H.}\ \bibnamefont {Pixley}},\ }\bibfield  {title} {\bibinfo {title} {{Critical properties of the measurement-induced transition in random quantum circuits}},\ }\href {https://doi.org/10.1103/PhysRevB.101.060301} {\bibfield  {journal} {\bibinfo  {journal} {Phys. Rev. B}\ }\textbf {\bibinfo {volume} {101}},\ \bibinfo {pages} {060301} (\bibinfo {year} {2020})}\BibitemShut {NoStop}%
\bibitem [{\citenamefont {Roser}\ \emph {et~al.}(2023)\citenamefont {Roser}, \citenamefont {B\"uchler},\ and\ \citenamefont {Lang}}]{Roser2023}%
  \BibitemOpen
  \bibfield  {author} {\bibinfo {author} {\bibfnamefont {F.}~\bibnamefont {Roser}}, \bibinfo {author} {\bibfnamefont {H.~P.}\ \bibnamefont {B\"uchler}},\ and\ \bibinfo {author} {\bibfnamefont {N.}~\bibnamefont {Lang}},\ }\bibfield  {title} {\bibinfo {title} {{Decoding the projective transverse field Ising model}},\ }\href {https://doi.org/10.1103/PhysRevB.107.214201} {\bibfield  {journal} {\bibinfo  {journal} {Phys. Rev. B}\ }\textbf {\bibinfo {volume} {107}},\ \bibinfo {pages} {214201} (\bibinfo {year} {2023})}\BibitemShut {NoStop}%
\bibitem [{\citenamefont {Roser}\ \emph {et~al.}(2026)\citenamefont {Roser}, \citenamefont {Springer}, \citenamefont {B\"uchler},\ and\ \citenamefont {Lang}}]{Roser2026}%
  \BibitemOpen
  \bibfield  {author} {\bibinfo {author} {\bibfnamefont {F.}~\bibnamefont {Roser}}, \bibinfo {author} {\bibfnamefont {E.~M.}\ \bibnamefont {Springer}}, \bibinfo {author} {\bibfnamefont {H.~P.}\ \bibnamefont {B\"uchler}},\ and\ \bibinfo {author} {\bibfnamefont {N.}~\bibnamefont {Lang}},\ }\bibfield  {title} {\bibinfo {title} {{Robust Detection of an Entanglement Transition in the Projective Transverse-Field Ising Model}},\ }\href {https://doi.org/10.1103/5m5q-f7vc} {\bibfield  {journal} {\bibinfo  {journal} {Phys. Rev. Lett.}\ }\textbf {\bibinfo {volume} {136}},\ \bibinfo {pages} {140403} (\bibinfo {year} {2026})}\BibitemShut {NoStop}%
\bibitem [{\citenamefont {Ippoliti}\ \emph {et~al.}(2021)\citenamefont {Ippoliti}, \citenamefont {Gullans}, \citenamefont {Gopalakrishnan}, \citenamefont {Huse},\ and\ \citenamefont {Khemani}}]{Vedika2021measure}%
  \BibitemOpen
  \bibfield  {author} {\bibinfo {author} {\bibfnamefont {M.}~\bibnamefont {Ippoliti}}, \bibinfo {author} {\bibfnamefont {M.~J.}\ \bibnamefont {Gullans}}, \bibinfo {author} {\bibfnamefont {S.}~\bibnamefont {Gopalakrishnan}}, \bibinfo {author} {\bibfnamefont {D.~A.}\ \bibnamefont {Huse}},\ and\ \bibinfo {author} {\bibfnamefont {V.}~\bibnamefont {Khemani}},\ }\bibfield  {title} {\bibinfo {title} {{Entanglement Phase Transitions in Measurement-Only Dynamics}},\ }\href {https://doi.org/10.1103/PhysRevX.11.011030} {\bibfield  {journal} {\bibinfo  {journal} {Phys. Rev. X}\ }\textbf {\bibinfo {volume} {11}},\ \bibinfo {pages} {011030} (\bibinfo {year} {2021})}\BibitemShut {NoStop}%
\bibitem [{\citenamefont {{Lavasani}}\ \emph {et~al.}(2021)\citenamefont {{Lavasani}}, \citenamefont {{Alavirad}},\ and\ \citenamefont {{Barkeshli}}}]{Barkeshli2021measure}%
  \BibitemOpen
  \bibfield  {author} {\bibinfo {author} {\bibfnamefont {A.}~\bibnamefont {{Lavasani}}}, \bibinfo {author} {\bibfnamefont {Y.}~\bibnamefont {{Alavirad}}},\ and\ \bibinfo {author} {\bibfnamefont {M.}~\bibnamefont {{Barkeshli}}},\ }\bibfield  {title} {\bibinfo {title} {{Measurement-induced topological entanglement transitions in symmetric random quantum circuits}},\ }\href {https://doi.org/10.1038/s41567-020-01112-z} {\bibfield  {journal} {\bibinfo  {journal} {Nat. Phys.}\ }\textbf {\bibinfo {volume} {17}},\ \bibinfo {pages} {342} (\bibinfo {year} {2021})}\BibitemShut {NoStop}%
\bibitem [{\citenamefont {Nahum}\ \emph {et~al.}(2021)\citenamefont {Nahum}, \citenamefont {Roy}, \citenamefont {Skinner},\ and\ \citenamefont {Ruhman}}]{Nahum2021forcedmeasurement}%
  \BibitemOpen
  \bibfield  {author} {\bibinfo {author} {\bibfnamefont {A.}~\bibnamefont {Nahum}}, \bibinfo {author} {\bibfnamefont {S.}~\bibnamefont {Roy}}, \bibinfo {author} {\bibfnamefont {B.}~\bibnamefont {Skinner}},\ and\ \bibinfo {author} {\bibfnamefont {J.}~\bibnamefont {Ruhman}},\ }\bibfield  {title} {\bibinfo {title} {{Measurement and Entanglement Phase Transitions in All-To-All Quantum Circuits, on Quantum Trees, and in Landau-Ginsburg Theory}},\ }\href {https://doi.org/10.1103/PRXQuantum.2.010352} {\bibfield  {journal} {\bibinfo  {journal} {PRX Quantum}\ }\textbf {\bibinfo {volume} {2}},\ \bibinfo {pages} {010352} (\bibinfo {year} {2021})}\BibitemShut {NoStop}%
\bibitem [{\citenamefont {Zhu}\ \emph {et~al.}(2024)\citenamefont {Zhu}, \citenamefont {Tantivasadakarn},\ and\ \citenamefont {Trebst}}]{Zhu23structuredVolumeLaw}%
  \BibitemOpen
  \bibfield  {author} {\bibinfo {author} {\bibfnamefont {G.-Y.}\ \bibnamefont {Zhu}}, \bibinfo {author} {\bibfnamefont {N.}~\bibnamefont {Tantivasadakarn}},\ and\ \bibinfo {author} {\bibfnamefont {S.}~\bibnamefont {Trebst}},\ }\bibfield  {title} {\bibinfo {title} {{Structured volume-law entanglement in an interacting, monitored Majorana spin liquid}},\ }\href {https://doi.org/10.1103/PhysRevResearch.6.L042063} {\bibfield  {journal} {\bibinfo  {journal} {Phys. Rev. Res.}\ }\textbf {\bibinfo {volume} {6}},\ \bibinfo {pages} {L042063} (\bibinfo {year} {2024})}\BibitemShut {NoStop}%
\bibitem [{\citenamefont {Klocke}\ \emph {et~al.}(2025)\citenamefont {Klocke}, \citenamefont {Simm}, \citenamefont {Zhu}, \citenamefont {Trebst},\ and\ \citenamefont {Buchhold}}]{klocke2024entanglement}%
  \BibitemOpen
  \bibfield  {author} {\bibinfo {author} {\bibfnamefont {K.}~\bibnamefont {Klocke}}, \bibinfo {author} {\bibfnamefont {D.}~\bibnamefont {Simm}}, \bibinfo {author} {\bibfnamefont {G.-Y.}\ \bibnamefont {Zhu}}, \bibinfo {author} {\bibfnamefont {S.}~\bibnamefont {Trebst}},\ and\ \bibinfo {author} {\bibfnamefont {M.}~\bibnamefont {Buchhold}},\ }\bibfield  {title} {\bibinfo {title} {{Entanglement dynamics in monitored Kitaev circuits: Loop models, symmetry classification, and quantum Lifshitz scaling}},\ }\href {https://doi.org/10.1103/PhysRevB.111.224301} {\bibfield  {journal} {\bibinfo  {journal} {Phys. Rev. B}\ }\textbf {\bibinfo {volume} {111}},\ \bibinfo {pages} {224301} (\bibinfo {year} {2025})}\BibitemShut {NoStop}%
\bibitem [{\citenamefont {Nahum}\ and\ \citenamefont {Skinner}(2020)}]{Nahum20freefermion}%
  \BibitemOpen
  \bibfield  {author} {\bibinfo {author} {\bibfnamefont {A.}~\bibnamefont {Nahum}}\ and\ \bibinfo {author} {\bibfnamefont {B.}~\bibnamefont {Skinner}},\ }\bibfield  {title} {\bibinfo {title} {{Entanglement and dynamics of diffusion-annihilation processes with Majorana defects}},\ }\href {https://doi.org/10.1103/PhysRevResearch.2.023288} {\bibfield  {journal} {\bibinfo  {journal} {Phys. Rev. Res.}\ }\textbf {\bibinfo {volume} {2}},\ \bibinfo {pages} {023288} (\bibinfo {year} {2020})}\BibitemShut {NoStop}%
\bibitem [{\citenamefont {Kells}\ \emph {et~al.}(2023)\citenamefont {Kells}, \citenamefont {Meidan},\ and\ \citenamefont {Romito}}]{Graham23majorana}%
  \BibitemOpen
  \bibfield  {author} {\bibinfo {author} {\bibfnamefont {G.}~\bibnamefont {Kells}}, \bibinfo {author} {\bibfnamefont {D.}~\bibnamefont {Meidan}},\ and\ \bibinfo {author} {\bibfnamefont {A.}~\bibnamefont {Romito}},\ }\bibfield  {title} {\bibinfo {title} {{Topological transitions in weakly monitored free fermions}},\ }\href {https://doi.org/10.21468/SciPostPhys.14.3.031} {\bibfield  {journal} {\bibinfo  {journal} {SciPost Phys.}\ }\textbf {\bibinfo {volume} {14}},\ \bibinfo {pages} {031} (\bibinfo {year} {2023})}\BibitemShut {NoStop}%
\bibitem [{\citenamefont {Kitaev}(2001)}]{Kitaev2001}%
  \BibitemOpen
  \bibfield  {author} {\bibinfo {author} {\bibfnamefont {A.~Y.}\ \bibnamefont {Kitaev}},\ }\bibfield  {title} {\bibinfo {title} {{Unpaired Majorana fermions in quantum wires}},\ }\href {https://doi.org/10.1070/1063-7869/44/10S/S29} {\bibfield  {journal} {\bibinfo  {journal} {Physics-Uspekhi}\ }\textbf {\bibinfo {volume} {44}},\ \bibinfo {pages} {131} (\bibinfo {year} {2001})}\BibitemShut {NoStop}%
\bibitem [{\citenamefont {Dumitrescu}\ \emph {et~al.}(2018)\citenamefont {Dumitrescu}, \citenamefont {Vasseur},\ and\ \citenamefont {Potter}}]{Dumitrescu2018}%
  \BibitemOpen
  \bibfield  {author} {\bibinfo {author} {\bibfnamefont {P.~T.}\ \bibnamefont {Dumitrescu}}, \bibinfo {author} {\bibfnamefont {R.}~\bibnamefont {Vasseur}},\ and\ \bibinfo {author} {\bibfnamefont {A.~C.}\ \bibnamefont {Potter}},\ }\bibfield  {title} {\bibinfo {title} {{Logarithmically Slow Relaxation in Quasiperiodically Driven Random Spin Chains}},\ }\href {https://doi.org/10.1103/PhysRevLett.120.070602} {\bibfield  {journal} {\bibinfo  {journal} {Phys. Rev. Lett.}\ }\textbf {\bibinfo {volume} {120}},\ \bibinfo {pages} {070602} (\bibinfo {year} {2018})}\BibitemShut {NoStop}%
\bibitem [{\citenamefont {Maity}\ \emph {et~al.}(2019)\citenamefont {Maity}, \citenamefont {Bhattacharya}, \citenamefont {Dutta},\ and\ \citenamefont {Sen}}]{Maity2019}%
  \BibitemOpen
  \bibfield  {author} {\bibinfo {author} {\bibfnamefont {S.}~\bibnamefont {Maity}}, \bibinfo {author} {\bibfnamefont {U.}~\bibnamefont {Bhattacharya}}, \bibinfo {author} {\bibfnamefont {A.}~\bibnamefont {Dutta}},\ and\ \bibinfo {author} {\bibfnamefont {D.}~\bibnamefont {Sen}},\ }\bibfield  {title} {\bibinfo {title} {{Fibonacci steady states in a driven integrable quantum system}},\ }\href {https://doi.org/10.1103/PhysRevB.99.020306} {\bibfield  {journal} {\bibinfo  {journal} {Phys. Rev. B}\ }\textbf {\bibinfo {volume} {99}},\ \bibinfo {pages} {020306} (\bibinfo {year} {2019})}\BibitemShut {NoStop}%
\bibitem [{\citenamefont {Lapierre}\ \emph {et~al.}(2020)\citenamefont {Lapierre}, \citenamefont {Choo}, \citenamefont {Tiwari}, \citenamefont {Tauber}, \citenamefont {Neupert},\ and\ \citenamefont {Chitra}}]{Lapierre2020}%
  \BibitemOpen
  \bibfield  {author} {\bibinfo {author} {\bibfnamefont {B.}~\bibnamefont {Lapierre}}, \bibinfo {author} {\bibfnamefont {K.}~\bibnamefont {Choo}}, \bibinfo {author} {\bibfnamefont {A.}~\bibnamefont {Tiwari}}, \bibinfo {author} {\bibfnamefont {C.}~\bibnamefont {Tauber}}, \bibinfo {author} {\bibfnamefont {T.}~\bibnamefont {Neupert}},\ and\ \bibinfo {author} {\bibfnamefont {R.}~\bibnamefont {Chitra}},\ }\bibfield  {title} {\bibinfo {title} {{Fine structure of heating in a quasiperiodically driven critical quantum system}},\ }\href {https://doi.org/10.1103/PhysRevResearch.2.033461} {\bibfield  {journal} {\bibinfo  {journal} {Phys. Rev. Res.}\ }\textbf {\bibinfo {volume} {2}},\ \bibinfo {pages} {033461} (\bibinfo {year} {2020})}\BibitemShut {NoStop}%
\bibitem [{\citenamefont {Wen}\ \emph {et~al.}(2021)\citenamefont {Wen}, \citenamefont {Fan}, \citenamefont {Vishwanath},\ and\ \citenamefont {Gu}}]{Wen_2021}%
  \BibitemOpen
  \bibfield  {author} {\bibinfo {author} {\bibfnamefont {X.}~\bibnamefont {Wen}}, \bibinfo {author} {\bibfnamefont {R.}~\bibnamefont {Fan}}, \bibinfo {author} {\bibfnamefont {A.}~\bibnamefont {Vishwanath}},\ and\ \bibinfo {author} {\bibfnamefont {Y.}~\bibnamefont {Gu}},\ }\bibfield  {title} {\bibinfo {title} {{Periodically, quasiperiodically, and randomly driven conformal field theories}},\ }\href {https://doi.org/10.1103/PhysRevResearch.3.023044} {\bibfield  {journal} {\bibinfo  {journal} {Phys. Rev. Res.}\ }\textbf {\bibinfo {volume} {3}},\ \bibinfo {pages} {023044} (\bibinfo {year} {2021})}\BibitemShut {NoStop}%
\bibitem [{\citenamefont {Bhattacharjee}\ \emph {et~al.}(2022)\citenamefont {Bhattacharjee}, \citenamefont {Bandyopadhyay},\ and\ \citenamefont {Dutta}}]{Bhattacharjee2022}%
  \BibitemOpen
  \bibfield  {author} {\bibinfo {author} {\bibfnamefont {S.}~\bibnamefont {Bhattacharjee}}, \bibinfo {author} {\bibfnamefont {S.}~\bibnamefont {Bandyopadhyay}},\ and\ \bibinfo {author} {\bibfnamefont {A.}~\bibnamefont {Dutta}},\ }\bibfield  {title} {\bibinfo {title} {{Quasilocalization dynamics in a Fibonacci quantum rotor}},\ }\href {https://doi.org/10.1103/PhysRevA.106.022206} {\bibfield  {journal} {\bibinfo  {journal} {Phys. Rev. A}\ }\textbf {\bibinfo {volume} {106}},\ \bibinfo {pages} {022206} (\bibinfo {year} {2022})}\BibitemShut {NoStop}%
\bibitem [{\citenamefont {Schmid}\ \emph {et~al.}(2025)\citenamefont {Schmid}, \citenamefont {Peng}, \citenamefont {Refael},\ and\ \citenamefont {von Oppen}}]{schmid24fibonacci}%
  \BibitemOpen
  \bibfield  {author} {\bibinfo {author} {\bibfnamefont {H.}~\bibnamefont {Schmid}}, \bibinfo {author} {\bibfnamefont {Y.}~\bibnamefont {Peng}}, \bibinfo {author} {\bibfnamefont {G.}~\bibnamefont {Refael}},\ and\ \bibinfo {author} {\bibfnamefont {F.}~\bibnamefont {von Oppen}},\ }\bibfield  {title} {\bibinfo {title} {{Self-Similar Phase Diagram of the Fibonacci-Driven Quantum Ising Model}},\ }\href {https://doi.org/10.1103/hn66-j8pt} {\bibfield  {journal} {\bibinfo  {journal} {Phys. Rev. Lett.}\ }\textbf {\bibinfo {volume} {134}},\ \bibinfo {pages} {240404} (\bibinfo {year} {2025})}\BibitemShut {NoStop}%
\bibitem [{\citenamefont {Lapierre}\ \emph {et~al.}(2025)\citenamefont {Lapierre}, \citenamefont {Mo},\ and\ \citenamefont {Ryu}}]{Lapierre2025}%
  \BibitemOpen
  \bibfield  {author} {\bibinfo {author} {\bibfnamefont {B.}~\bibnamefont {Lapierre}}, \bibinfo {author} {\bibfnamefont {L.-H.}\ \bibnamefont {Mo}},\ and\ \bibinfo {author} {\bibfnamefont {S.}~\bibnamefont {Ryu}},\ }\href {https://arxiv.org/abs/2507.03768} {\bibinfo {title} {{Entanglement transitions in structured and random nonunitary Gaussian circuits}}} (\bibinfo {year} {2025}),\ \Eprint {https://arxiv.org/abs/2507.03768} {arXiv:2507.03768 [quant-ph]} \BibitemShut {NoStop}%
\bibitem [{\citenamefont {Kohmoto}\ \emph {et~al.}(1983)\citenamefont {Kohmoto}, \citenamefont {Kadanoff},\ and\ \citenamefont {Tang}}]{Kohmoto1983qpising}%
  \BibitemOpen
  \bibfield  {author} {\bibinfo {author} {\bibfnamefont {M.}~\bibnamefont {Kohmoto}}, \bibinfo {author} {\bibfnamefont {L.~P.}\ \bibnamefont {Kadanoff}},\ and\ \bibinfo {author} {\bibfnamefont {C.}~\bibnamefont {Tang}},\ }\bibfield  {title} {\bibinfo {title} {{Localization Problem in One Dimension: Mapping and Escape}},\ }\href {https://doi.org/10.1103/PhysRevLett.50.1870} {\bibfield  {journal} {\bibinfo  {journal} {Phys. Rev. Lett.}\ }\textbf {\bibinfo {volume} {50}},\ \bibinfo {pages} {1870} (\bibinfo {year} {1983})}\BibitemShut {NoStop}%
\bibitem [{\citenamefont {Ostlund}\ \emph {et~al.}(1983)\citenamefont {Ostlund}, \citenamefont {Pandit}, \citenamefont {Rand}, \citenamefont {Schellnhuber},\ and\ \citenamefont {Siggia}}]{Siggia83qpising}%
  \BibitemOpen
  \bibfield  {author} {\bibinfo {author} {\bibfnamefont {S.}~\bibnamefont {Ostlund}}, \bibinfo {author} {\bibfnamefont {R.}~\bibnamefont {Pandit}}, \bibinfo {author} {\bibfnamefont {D.}~\bibnamefont {Rand}}, \bibinfo {author} {\bibfnamefont {H.~J.}\ \bibnamefont {Schellnhuber}},\ and\ \bibinfo {author} {\bibfnamefont {E.~D.}\ \bibnamefont {Siggia}},\ }\bibfield  {title} {\bibinfo {title} {{One-Dimensional Schr\"odinger Equation with an Almost Periodic Potential}},\ }\href {https://doi.org/10.1103/PhysRevLett.50.1873} {\bibfield  {journal} {\bibinfo  {journal} {Phys. Rev. Lett.}\ }\textbf {\bibinfo {volume} {50}},\ \bibinfo {pages} {1873} (\bibinfo {year} {1983})}\BibitemShut {NoStop}%
\bibitem [{\citenamefont {Chandran}\ and\ \citenamefont {Laumann}(2017)}]{Chandran17qpising}%
  \BibitemOpen
  \bibfield  {author} {\bibinfo {author} {\bibfnamefont {A.}~\bibnamefont {Chandran}}\ and\ \bibinfo {author} {\bibfnamefont {C.~R.}\ \bibnamefont {Laumann}},\ }\bibfield  {title} {\bibinfo {title} {{Localization and Symmetry Breaking in the Quantum Quasiperiodic Ising Glass}},\ }\href {https://doi.org/10.1103/PhysRevX.7.031061} {\bibfield  {journal} {\bibinfo  {journal} {Phys. Rev. X}\ }\textbf {\bibinfo {volume} {7}},\ \bibinfo {pages} {031061} (\bibinfo {year} {2017})}\BibitemShut {NoStop}%
\bibitem [{\citenamefont {Crowley}\ \emph {et~al.}(2022)\citenamefont {Crowley}, \citenamefont {Laumann},\ and\ \citenamefont {Chandran}}]{Crowley18qpising}%
  \BibitemOpen
  \bibfield  {author} {\bibinfo {author} {\bibfnamefont {P.~J.~D.}\ \bibnamefont {Crowley}}, \bibinfo {author} {\bibfnamefont {C.~R.}\ \bibnamefont {Laumann}},\ and\ \bibinfo {author} {\bibfnamefont {A.}~\bibnamefont {Chandran}},\ }\bibfield  {title} {\bibinfo {title} {{Critical behaviour of the quasi-periodic quantum Ising chain}},\ }\href {https://doi.org/10.1088/1742-5468/ac815d} {\bibfield  {journal} {\bibinfo  {journal} {J. Stat. Mech.: Theory Exp.}\ }\textbf {\bibinfo {volume} {2022}}\bibinfo  {number} { (8)},\ \bibinfo {pages} {083102}}\BibitemShut {NoStop}%
\bibitem [{\citenamefont {Peng}\ and\ \citenamefont {Refael}(2018)}]{PengRefael18}%
  \BibitemOpen
\bibfield  {number} {  }\bibfield  {author} {\bibinfo {author} {\bibfnamefont {Y.}~\bibnamefont {Peng}}\ and\ \bibinfo {author} {\bibfnamefont {G.}~\bibnamefont {Refael}},\ }\bibfield  {title} {\bibinfo {title} {{Time-quasiperiodic topological superconductors with Majorana multiplexing}},\ }\href {https://doi.org/10.1103/PhysRevB.98.220509} {\bibfield  {journal} {\bibinfo  {journal} {Phys. Rev. B}\ }\textbf {\bibinfo {volume} {98}},\ \bibinfo {pages} {220509} (\bibinfo {year} {2018})}\BibitemShut {NoStop}%
\bibitem [{\citenamefont {Else}\ \emph {et~al.}(2020)\citenamefont {Else}, \citenamefont {Ho},\ and\ \citenamefont {Dumitrescu}}]{ElseHo20quasiperiodic}%
  \BibitemOpen
  \bibfield  {author} {\bibinfo {author} {\bibfnamefont {D.~V.}\ \bibnamefont {Else}}, \bibinfo {author} {\bibfnamefont {W.~W.}\ \bibnamefont {Ho}},\ and\ \bibinfo {author} {\bibfnamefont {P.~T.}\ \bibnamefont {Dumitrescu}},\ }\bibfield  {title} {\bibinfo {title} {{Long-Lived Interacting Phases of Matter Protected by Multiple Time-Translation Symmetries in Quasiperiodically Driven Systems}},\ }\href {https://doi.org/10.1103/PhysRevX.10.021032} {\bibfield  {journal} {\bibinfo  {journal} {Phys. Rev. X}\ }\textbf {\bibinfo {volume} {10}},\ \bibinfo {pages} {021032} (\bibinfo {year} {2020})}\BibitemShut {NoStop}%
\bibitem [{\citenamefont {Garratt}\ \emph {et~al.}(2023)\citenamefont {Garratt}, \citenamefont {Weinstein},\ and\ \citenamefont {Altman}}]{Garratt22}%
  \BibitemOpen
  \bibfield  {author} {\bibinfo {author} {\bibfnamefont {S.~J.}\ \bibnamefont {Garratt}}, \bibinfo {author} {\bibfnamefont {Z.}~\bibnamefont {Weinstein}},\ and\ \bibinfo {author} {\bibfnamefont {E.}~\bibnamefont {Altman}},\ }\bibfield  {title} {\bibinfo {title} {{Measurements Conspire Nonlocally to Restructure Critical Quantum States}},\ }\href {https://doi.org/10.1103/PhysRevX.13.021026} {\bibfield  {journal} {\bibinfo  {journal} {Phys. Rev. X}\ }\textbf {\bibinfo {volume} {13}},\ \bibinfo {pages} {021026} (\bibinfo {year} {2023})}\BibitemShut {NoStop}%
\bibitem [{\citenamefont {Weinstein}\ \emph {et~al.}(2023)\citenamefont {Weinstein}, \citenamefont {Sajith}, \citenamefont {Altman},\ and\ \citenamefont {Garratt}}]{Garratt23measureising}%
  \BibitemOpen
  \bibfield  {author} {\bibinfo {author} {\bibfnamefont {Z.}~\bibnamefont {Weinstein}}, \bibinfo {author} {\bibfnamefont {R.}~\bibnamefont {Sajith}}, \bibinfo {author} {\bibfnamefont {E.}~\bibnamefont {Altman}},\ and\ \bibinfo {author} {\bibfnamefont {S.~J.}\ \bibnamefont {Garratt}},\ }\bibfield  {title} {\bibinfo {title} {{Nonlocality and entanglement in measured critical quantum Ising chains}},\ }\href {https://doi.org/10.1103/PhysRevB.107.245132} {\bibfield  {journal} {\bibinfo  {journal} {Phys. Rev. B}\ }\textbf {\bibinfo {volume} {107}},\ \bibinfo {pages} {245132} (\bibinfo {year} {2023})}\BibitemShut {NoStop}%
\bibitem [{\citenamefont {Murciano}\ \emph {et~al.}(2023)\citenamefont {Murciano}, \citenamefont {Sala}, \citenamefont {Liu}, \citenamefont {Mong},\ and\ \citenamefont {Alicea}}]{Alicea23measureising}%
  \BibitemOpen
  \bibfield  {author} {\bibinfo {author} {\bibfnamefont {S.}~\bibnamefont {Murciano}}, \bibinfo {author} {\bibfnamefont {P.}~\bibnamefont {Sala}}, \bibinfo {author} {\bibfnamefont {Y.}~\bibnamefont {Liu}}, \bibinfo {author} {\bibfnamefont {R.~S.~K.}\ \bibnamefont {Mong}},\ and\ \bibinfo {author} {\bibfnamefont {J.}~\bibnamefont {Alicea}},\ }\bibfield  {title} {\bibinfo {title} {{Measurement-Altered Ising Quantum Criticality}},\ }\href {https://doi.org/10.1103/PhysRevX.13.041042} {\bibfield  {journal} {\bibinfo  {journal} {Phys. Rev. X}\ }\textbf {\bibinfo {volume} {13}},\ \bibinfo {pages} {041042} (\bibinfo {year} {2023})}\BibitemShut {NoStop}%
\bibitem [{\citenamefont {Yang}\ \emph {et~al.}(2023)\citenamefont {Yang}, \citenamefont {Mao},\ and\ \citenamefont {Jian}}]{jian23measureising}%
  \BibitemOpen
  \bibfield  {author} {\bibinfo {author} {\bibfnamefont {Z.}~\bibnamefont {Yang}}, \bibinfo {author} {\bibfnamefont {D.}~\bibnamefont {Mao}},\ and\ \bibinfo {author} {\bibfnamefont {C.-M.}\ \bibnamefont {Jian}},\ }\bibfield  {title} {\bibinfo {title} {{Entanglement in a one-dimensional critical state after measurements}},\ }\href {https://doi.org/10.1103/PhysRevB.108.165120} {\bibfield  {journal} {\bibinfo  {journal} {Phys. Rev. B}\ }\textbf {\bibinfo {volume} {108}},\ \bibinfo {pages} {165120} (\bibinfo {year} {2023})}\BibitemShut {NoStop}%
\bibitem [{\citenamefont {Sun}\ \emph {et~al.}(2023)\citenamefont {Sun}, \citenamefont {Yao},\ and\ \citenamefont {Jian}}]{Sun2023}%
  \BibitemOpen
  \bibfield  {author} {\bibinfo {author} {\bibfnamefont {X.}~\bibnamefont {Sun}}, \bibinfo {author} {\bibfnamefont {H.}~\bibnamefont {Yao}},\ and\ \bibinfo {author} {\bibfnamefont {S.-K.}\ \bibnamefont {Jian}},\ }\href {https://arxiv.org/abs/2301.11337} {\bibinfo {title} {{New critical states induced by measurement}}} (\bibinfo {year} {2023}),\ \Eprint {https://arxiv.org/abs/2301.11337} {arXiv:2301.11337 [quant-ph]} \BibitemShut {NoStop}%
\bibitem [{\citenamefont {Paviglianiti}\ \emph {et~al.}(2024)\citenamefont {Paviglianiti}, \citenamefont {Turkeshi}, \citenamefont {Schir{\`{o}}},\ and\ \citenamefont {Silva}}]{Paviglianiti2024}%
  \BibitemOpen
  \bibfield  {author} {\bibinfo {author} {\bibfnamefont {A.}~\bibnamefont {Paviglianiti}}, \bibinfo {author} {\bibfnamefont {X.}~\bibnamefont {Turkeshi}}, \bibinfo {author} {\bibfnamefont {M.}~\bibnamefont {Schir{\`{o}}}},\ and\ \bibinfo {author} {\bibfnamefont {A.}~\bibnamefont {Silva}},\ }\bibfield  {title} {\bibinfo {title} {{Enhanced {E}ntanglement in the {M}easurement-{A}ltered {Q}uantum {I}sing {C}hain}},\ }\href {https://doi.org/10.22331/q-2024-12-23-1576} {\bibfield  {journal} {\bibinfo  {journal} {{Quantum}}\ }\textbf {\bibinfo {volume} {8}},\ \bibinfo {pages} {1576} (\bibinfo {year} {2024})}\BibitemShut {NoStop}%
\bibitem [{Note1()}]{Note1}%
  \BibitemOpen
  \bibinfo {note} {Such an initialization allows for the formation of LRE states in the SSB phase. More concretely, $\protect \mathbb {Z}_2$ symmetry is preserved through the circuit, and thus the late-time state in the SSB phase corresponds to a Greenberger-Horne-Zeilinger type cat state, with long-range entanglement.}\BibitemShut {Stop}%
\bibitem [{\citenamefont {{Gottesman}}(1998)}]{GottesmanKnill98}%
  \BibitemOpen
  \bibfield  {author} {\bibinfo {author} {\bibfnamefont {D.}~\bibnamefont {{Gottesman}}},\ }\bibfield  {title} {\bibinfo {title} {{The Heisenberg Representation of Quantum Computers}},\ }\href@noop {} {\bibfield  {journal} {\bibinfo  {journal} {preprint}\ } (\bibinfo {year} {1998})},\ \Eprint {https://arxiv.org/abs/quant-ph/9807006} {arXiv:quant-ph/9807006} \BibitemShut {NoStop}%
\bibitem [{\citenamefont {Aaronson}\ and\ \citenamefont {Gottesman}(2004)}]{AaronsonGottesman04}%
  \BibitemOpen
  \bibfield  {author} {\bibinfo {author} {\bibfnamefont {S.}~\bibnamefont {Aaronson}}\ and\ \bibinfo {author} {\bibfnamefont {D.}~\bibnamefont {Gottesman}},\ }\bibfield  {title} {\bibinfo {title} {{Improved simulation of stabilizer circuits}},\ }\href {https://doi.org/10.1103/PhysRevA.70.052328} {\bibfield  {journal} {\bibinfo  {journal} {Phys. Rev. A}\ }\textbf {\bibinfo {volume} {70}},\ \bibinfo {pages} {052328} (\bibinfo {year} {2004})}\BibitemShut {NoStop}%
\bibitem [{Note2()}]{Note2}%
  \BibitemOpen
  \bibinfo {note} {For a more precise determination of the critical locations we also adopt the ``coherent information'' as a diagnostic, which shows the best finite-size scaling data collapse performance~\cite {Wang25selfdual,Wan25nishimori, Eckstein25learning}}\BibitemShut {NoStop}%
\bibitem [{Note3()}]{Note3}%
  \BibitemOpen
  \bibinfo {note} {Note that while the cat state maintains $\protect \mathbb {Z}_2$ symmetry microscopically, in terms of phase of matter we still refer to it as $\protect \mathbb {Z}_2$ SSB phase, defined by the long-range correlation}\BibitemShut {NoStop}%
\bibitem [{\citenamefont {Buča}\ and\ \citenamefont {Prosen}(2012)}]{Prosen12weaksym}%
  \BibitemOpen
  \bibfield  {author} {\bibinfo {author} {\bibfnamefont {B.}~\bibnamefont {Buča}}\ and\ \bibinfo {author} {\bibfnamefont {T.}~\bibnamefont {Prosen}},\ }\bibfield  {title} {\bibinfo {title} {{A note on symmetry reductions of the Lindblad equation: transport in constrained open spin chains}},\ }\href {https://doi.org/10.1088/1367-2630/14/7/073007} {\bibfield  {journal} {\bibinfo  {journal} {New J. Phys.}\ }\textbf {\bibinfo {volume} {14}},\ \bibinfo {pages} {073007} (\bibinfo {year} {2012})}\BibitemShut {NoStop}%
\bibitem [{\citenamefont {Albert}\ and\ \citenamefont {Jiang}(2014)}]{Jiang14weaksym}%
  \BibitemOpen
  \bibfield  {author} {\bibinfo {author} {\bibfnamefont {V.~V.}\ \bibnamefont {Albert}}\ and\ \bibinfo {author} {\bibfnamefont {L.}~\bibnamefont {Jiang}},\ }\bibfield  {title} {\bibinfo {title} {{Symmetries and conserved quantities in Lindblad master equations}},\ }\href {https://doi.org/10.1103/PhysRevA.89.022118} {\bibfield  {journal} {\bibinfo  {journal} {Phys. Rev. A}\ }\textbf {\bibinfo {volume} {89}},\ \bibinfo {pages} {022118} (\bibinfo {year} {2014})}\BibitemShut {NoStop}%
\bibitem [{\citenamefont {Lieu}\ \emph {et~al.}(2020)\citenamefont {Lieu}, \citenamefont {Belyansky}, \citenamefont {Young}, \citenamefont {Lundgren}, \citenamefont {Albert},\ and\ \citenamefont {Gorshkov}}]{Gorshkov20weaksym}%
  \BibitemOpen
  \bibfield  {author} {\bibinfo {author} {\bibfnamefont {S.}~\bibnamefont {Lieu}}, \bibinfo {author} {\bibfnamefont {R.}~\bibnamefont {Belyansky}}, \bibinfo {author} {\bibfnamefont {J.~T.}\ \bibnamefont {Young}}, \bibinfo {author} {\bibfnamefont {R.}~\bibnamefont {Lundgren}}, \bibinfo {author} {\bibfnamefont {V.~V.}\ \bibnamefont {Albert}},\ and\ \bibinfo {author} {\bibfnamefont {A.~V.}\ \bibnamefont {Gorshkov}},\ }\bibfield  {title} {\bibinfo {title} {{Symmetry Breaking and Error Correction in Open Quantum Systems}},\ }\href {https://doi.org/10.1103/PhysRevLett.125.240405} {\bibfield  {journal} {\bibinfo  {journal} {Phys. Rev. Lett.}\ }\textbf {\bibinfo {volume} {125}},\ \bibinfo {pages} {240405} (\bibinfo {year} {2020})}\BibitemShut {NoStop}%
\bibitem [{\citenamefont {Ma}\ and\ \citenamefont {Wang}(2023)}]{Wang23averspt}%
  \BibitemOpen
  \bibfield  {author} {\bibinfo {author} {\bibfnamefont {R.}~\bibnamefont {Ma}}\ and\ \bibinfo {author} {\bibfnamefont {C.}~\bibnamefont {Wang}},\ }\bibfield  {title} {\bibinfo {title} {{Average Symmetry-Protected Topological Phases}},\ }\href {https://doi.org/10.1103/PhysRevX.13.031016} {\bibfield  {journal} {\bibinfo  {journal} {Phys. Rev. X}\ }\textbf {\bibinfo {volume} {13}},\ \bibinfo {pages} {031016} (\bibinfo {year} {2023})}\BibitemShut {NoStop}%
\bibitem [{\citenamefont {Lee}\ \emph {et~al.}(2023)\citenamefont {Lee}, \citenamefont {Jian},\ and\ \citenamefont {Xu}}]{Lee23decoher}%
  \BibitemOpen
  \bibfield  {author} {\bibinfo {author} {\bibfnamefont {J.~Y.}\ \bibnamefont {Lee}}, \bibinfo {author} {\bibfnamefont {C.-M.}\ \bibnamefont {Jian}},\ and\ \bibinfo {author} {\bibfnamefont {C.}~\bibnamefont {Xu}},\ }\bibfield  {title} {\bibinfo {title} {{Quantum Criticality Under Decoherence or Weak Measurement}},\ }\href {https://doi.org/10.1103/PRXQuantum.4.030317} {\bibfield  {journal} {\bibinfo  {journal} {PRX Quantum}\ }\textbf {\bibinfo {volume} {4}},\ \bibinfo {pages} {030317} (\bibinfo {year} {2023})}\BibitemShut {NoStop}%
\bibitem [{\citenamefont {Ma}\ \emph {et~al.}(2025)\citenamefont {Ma}, \citenamefont {Zhang}, \citenamefont {Bi}, \citenamefont {Cheng},\ and\ \citenamefont {Wang}}]{Wang23aversym}%
  \BibitemOpen
  \bibfield  {author} {\bibinfo {author} {\bibfnamefont {R.}~\bibnamefont {Ma}}, \bibinfo {author} {\bibfnamefont {J.-H.}\ \bibnamefont {Zhang}}, \bibinfo {author} {\bibfnamefont {Z.}~\bibnamefont {Bi}}, \bibinfo {author} {\bibfnamefont {M.}~\bibnamefont {Cheng}},\ and\ \bibinfo {author} {\bibfnamefont {C.}~\bibnamefont {Wang}},\ }\bibfield  {title} {\bibinfo {title} {{Topological Phases with Average Symmetries: The Decohered, the Disordered, and the Intrinsic}},\ }\href {https://doi.org/10.1103/PhysRevX.15.021062} {\bibfield  {journal} {\bibinfo  {journal} {Phys. Rev. X}\ }\textbf {\bibinfo {volume} {15}},\ \bibinfo {pages} {021062} (\bibinfo {year} {2025})}\BibitemShut {NoStop}%
\bibitem [{\citenamefont {Su}\ \emph {et~al.}(2024)\citenamefont {Su}, \citenamefont {Myerson-Jain}, \citenamefont {Wang}, \citenamefont {Jian},\ and\ \citenamefont {Xu}}]{Xu24higherformweaksym}%
  \BibitemOpen
  \bibfield  {author} {\bibinfo {author} {\bibfnamefont {K.}~\bibnamefont {Su}}, \bibinfo {author} {\bibfnamefont {N.}~\bibnamefont {Myerson-Jain}}, \bibinfo {author} {\bibfnamefont {C.}~\bibnamefont {Wang}}, \bibinfo {author} {\bibfnamefont {C.-M.}\ \bibnamefont {Jian}},\ and\ \bibinfo {author} {\bibfnamefont {C.}~\bibnamefont {Xu}},\ }\bibfield  {title} {\bibinfo {title} {{Higher-Form Symmetries under Weak Measurement}},\ }\href {https://doi.org/10.1103/PhysRevLett.132.200402} {\bibfield  {journal} {\bibinfo  {journal} {Phys. Rev. Lett.}\ }\textbf {\bibinfo {volume} {132}},\ \bibinfo {pages} {200402} (\bibinfo {year} {2024})}\BibitemShut {NoStop}%
\bibitem [{\citenamefont {P\"utz}\ \emph {et~al.}(2025)\citenamefont {P\"utz}, \citenamefont {Vasseur}, \citenamefont {Ludwig}, \citenamefont {Trebst},\ and\ \citenamefont {Zhu}}]{Puetz24}%
  \BibitemOpen
  \bibfield  {author} {\bibinfo {author} {\bibfnamefont {M.}~\bibnamefont {P\"utz}}, \bibinfo {author} {\bibfnamefont {R.}~\bibnamefont {Vasseur}}, \bibinfo {author} {\bibfnamefont {A.~W.}\ \bibnamefont {Ludwig}}, \bibinfo {author} {\bibfnamefont {S.}~\bibnamefont {Trebst}},\ and\ \bibinfo {author} {\bibfnamefont {G.-Y.}\ \bibnamefont {Zhu}},\ }\bibfield  {title} {\bibinfo {title} {{Flow to Nishimori Universality in Weakly Monitored Quantum Circuits with Qubit Loss}},\ }\href {https://doi.org/10.1103/ygfz-crvp} {\bibfield  {journal} {\bibinfo  {journal} {PRX Quantum}\ }\textbf {\bibinfo {volume} {6}},\ \bibinfo {pages} {040372} (\bibinfo {year} {2025})}\BibitemShut {NoStop}%
\bibitem [{Note4()}]{Note4}%
  \BibitemOpen
  \bibinfo {note} {Generally $c_{\protect \rm ent}$ is the scaling dimension of the boundary condition changing operator~\cite {Ludwig2020}. For the special case of unitary conformal field theory (such as for the post-selected trajectory), $c_{\protect \rm ent}$ is equivalent to the ``Casimir'' central charge determined via the free energy.}\BibitemShut {Stop}%
\bibitem [{\citenamefont {Calabrese}\ and\ \citenamefont {Cardy}(2004)}]{Calabrese2004}%
  \BibitemOpen
  \bibfield  {author} {\bibinfo {author} {\bibfnamefont {P.}~\bibnamefont {Calabrese}}\ and\ \bibinfo {author} {\bibfnamefont {J.}~\bibnamefont {Cardy}},\ }\bibfield  {title} {\bibinfo {title} {{Entanglement entropy and quantum field theory}},\ }\href {https://doi.org/10.1088/1742-5468/2004/06/P06002} {\bibfield  {journal} {\bibinfo  {journal} {J. Stat. Mech.}\ }\textbf {\bibinfo {volume} {2004}},\ \bibinfo {pages} {P06002} (\bibinfo {year} {2004})}\BibitemShut {NoStop}%
\bibitem [{\citenamefont {Calabrese}\ and\ \citenamefont {Cardy}(2009)}]{calabrese2009entanglement}%
  \BibitemOpen
  \bibfield  {author} {\bibinfo {author} {\bibfnamefont {P.}~\bibnamefont {Calabrese}}\ and\ \bibinfo {author} {\bibfnamefont {J.}~\bibnamefont {Cardy}},\ }\bibfield  {title} {\bibinfo {title} {{Entanglement entropy and conformal field theory}},\ }\href {https://doi.org/10.1088/1751-8113/42/50/504005} {\bibfield  {journal} {\bibinfo  {journal} {J. Phys. A: Math. Theor.}\ }\textbf {\bibinfo {volume} {42}},\ \bibinfo {pages} {504005} (\bibinfo {year} {2009})}\BibitemShut {NoStop}%
\bibitem [{\citenamefont {{Eckstein}}\ \emph {et~al.}(2025)\citenamefont {{Eckstein}}, \citenamefont {{Han}}, \citenamefont {{Trebst}},\ and\ \citenamefont {{Zhu}}}]{Eckstein25learning}%
  \BibitemOpen
  \bibfield  {author} {\bibinfo {author} {\bibfnamefont {F.}~\bibnamefont {{Eckstein}}}, \bibinfo {author} {\bibfnamefont {B.}~\bibnamefont {{Han}}}, \bibinfo {author} {\bibfnamefont {S.}~\bibnamefont {{Trebst}}},\ and\ \bibinfo {author} {\bibfnamefont {G.-Y.}\ \bibnamefont {{Zhu}}},\ }\href@noop {} {\bibinfo {title} {{Learning transitions of topological surface codes}}} (\bibinfo {year} {2025}),\ \Eprint {https://arxiv.org/abs/2512.19786} {arXiv:2512.19786} \BibitemShut {NoStop}%
\bibitem [{Note5()}]{Note5}%
  \BibitemOpen
  \bibinfo {note} {Notice that with our convention of the iterative rule $0\to 1,\ 1\to 10$, the ratio between the total number of ``1''s and ``0''s in a finite subsequence of the infinite Fibonacci word $10110101\protect \cdots $ is asymptotically given by $\phi =1.618...$}\BibitemShut {NoStop}%
\bibitem [{\citenamefont {Szyniszewski}\ \emph {et~al.}(2020)\citenamefont {Szyniszewski}, \citenamefont {Romito},\ and\ \citenamefont {Schomerus}}]{Schomerus2020weakmeasure}%
  \BibitemOpen
  \bibfield  {author} {\bibinfo {author} {\bibfnamefont {M.}~\bibnamefont {Szyniszewski}}, \bibinfo {author} {\bibfnamefont {A.}~\bibnamefont {Romito}},\ and\ \bibinfo {author} {\bibfnamefont {H.}~\bibnamefont {Schomerus}},\ }\bibfield  {title} {\bibinfo {title} {{Universality of Entanglement Transitions from Stroboscopic to Continuous Measurements}},\ }\href {https://doi.org/10.1103/PhysRevLett.125.210602} {\bibfield  {journal} {\bibinfo  {journal} {Phys. Rev. Lett.}\ }\textbf {\bibinfo {volume} {125}},\ \bibinfo {pages} {210602} (\bibinfo {year} {2020})}\BibitemShut {NoStop}%
\bibitem [{\citenamefont {Fuji}\ and\ \citenamefont {Ashida}(2020)}]{Ashida2020}%
  \BibitemOpen
  \bibfield  {author} {\bibinfo {author} {\bibfnamefont {Y.}~\bibnamefont {Fuji}}\ and\ \bibinfo {author} {\bibfnamefont {Y.}~\bibnamefont {Ashida}},\ }\bibfield  {title} {\bibinfo {title} {{Measurement-induced quantum criticality under continuous monitoring}},\ }\href {https://doi.org/10.1103/PhysRevB.102.054302} {\bibfield  {journal} {\bibinfo  {journal} {Phys. Rev. B}\ }\textbf {\bibinfo {volume} {102}},\ \bibinfo {pages} {054302} (\bibinfo {year} {2020})}\BibitemShut {NoStop}%
\bibitem [{\citenamefont {Fisher}(1992)}]{Fisher1992}%
  \BibitemOpen
  \bibfield  {author} {\bibinfo {author} {\bibfnamefont {D.~S.}\ \bibnamefont {Fisher}},\ }\bibfield  {title} {\bibinfo {title} {{Random transverse field Ising spin chains}},\ }\href {https://doi.org/10.1103/PhysRevLett.69.534} {\bibfield  {journal} {\bibinfo  {journal} {Phys. Rev. Lett.}\ }\textbf {\bibinfo {volume} {69}},\ \bibinfo {pages} {534} (\bibinfo {year} {1992})}\BibitemShut {NoStop}%
\bibitem [{\citenamefont {Gruzberg}\ \emph {et~al.}(2001)\citenamefont {Gruzberg}, \citenamefont {Read},\ and\ \citenamefont {Ludwig}}]{GRL2001}%
  \BibitemOpen
  \bibfield  {author} {\bibinfo {author} {\bibfnamefont {I.~A.}\ \bibnamefont {Gruzberg}}, \bibinfo {author} {\bibfnamefont {N.}~\bibnamefont {Read}},\ and\ \bibinfo {author} {\bibfnamefont {A.~W.~W.}\ \bibnamefont {Ludwig}},\ }\bibfield  {title} {\bibinfo {title} {{Random-bond Ising model in two dimensions: The Nishimori line and supersymmetry}},\ }\href {https://doi.org/10.1103/PhysRevB.63.104422} {\bibfield  {journal} {\bibinfo  {journal} {Phys. Rev. B}\ }\textbf {\bibinfo {volume} {63}},\ \bibinfo {pages} {104422} (\bibinfo {year} {2001})}\BibitemShut {NoStop}%
\bibitem [{\citenamefont {Levine}\ and\ \citenamefont {Steinhardt}(1986)}]{Levine1986}%
  \BibitemOpen
  \bibfield  {author} {\bibinfo {author} {\bibfnamefont {D.}~\bibnamefont {Levine}}\ and\ \bibinfo {author} {\bibfnamefont {P.~J.}\ \bibnamefont {Steinhardt}},\ }\bibfield  {title} {\bibinfo {title} {{Quasicrystals. I. Definition and structure}},\ }\href {https://doi.org/10.1103/PhysRevB.34.596} {\bibfield  {journal} {\bibinfo  {journal} {Phys. Rev. B}\ }\textbf {\bibinfo {volume} {34}},\ \bibinfo {pages} {596} (\bibinfo {year} {1986})}\BibitemShut {NoStop}%
\bibitem [{\citenamefont {D'Alessio}\ and\ \citenamefont {Rigol}(2014)}]{DAlessio2014}%
  \BibitemOpen
  \bibfield  {author} {\bibinfo {author} {\bibfnamefont {L.}~\bibnamefont {D'Alessio}}\ and\ \bibinfo {author} {\bibfnamefont {M.}~\bibnamefont {Rigol}},\ }\bibfield  {title} {\bibinfo {title} {{Long-time Behavior of Isolated Periodically Driven Interacting Lattice Systems}},\ }\href {https://doi.org/10.1103/PhysRevX.4.041048} {\bibfield  {journal} {\bibinfo  {journal} {Phys. Rev. X}\ }\textbf {\bibinfo {volume} {4}},\ \bibinfo {pages} {041048} (\bibinfo {year} {2014})}\BibitemShut {NoStop}%
\bibitem [{\citenamefont {Lazarides}\ \emph {et~al.}(2014)\citenamefont {Lazarides}, \citenamefont {Das},\ and\ \citenamefont {Moessner}}]{Lazarides2014}%
  \BibitemOpen
  \bibfield  {author} {\bibinfo {author} {\bibfnamefont {A.}~\bibnamefont {Lazarides}}, \bibinfo {author} {\bibfnamefont {A.}~\bibnamefont {Das}},\ and\ \bibinfo {author} {\bibfnamefont {R.}~\bibnamefont {Moessner}},\ }\bibfield  {title} {\bibinfo {title} {{Equilibrium states of generic quantum systems subject to periodic driving}},\ }\href {https://doi.org/10.1103/PhysRevE.90.012110} {\bibfield  {journal} {\bibinfo  {journal} {Phys. Rev. E}\ }\textbf {\bibinfo {volume} {90}},\ \bibinfo {pages} {012110} (\bibinfo {year} {2014})}\BibitemShut {NoStop}%
\bibitem [{\citenamefont {Abanin}\ \emph {et~al.}(2017)\citenamefont {Abanin}, \citenamefont {De~Roeck}, \citenamefont {Ho},\ and\ \citenamefont {Huveneers}}]{Abanin2017prethermal}%
  \BibitemOpen
  \bibfield  {author} {\bibinfo {author} {\bibfnamefont {D.~A.}\ \bibnamefont {Abanin}}, \bibinfo {author} {\bibfnamefont {W.}~\bibnamefont {De~Roeck}}, \bibinfo {author} {\bibfnamefont {W.~W.}\ \bibnamefont {Ho}},\ and\ \bibinfo {author} {\bibfnamefont {F.}~\bibnamefont {Huveneers}},\ }\bibfield  {title} {\bibinfo {title} {{Effective Hamiltonians, prethermalization, and slow energy absorption in periodically driven many-body systems}},\ }\href {https://doi.org/10.1103/PhysRevB.95.014112} {\bibfield  {journal} {\bibinfo  {journal} {Phys. Rev. B}\ }\textbf {\bibinfo {volume} {95}},\ \bibinfo {pages} {014112} (\bibinfo {year} {2017})}\BibitemShut {NoStop}%
\bibitem [{\citenamefont {Zhao}\ \emph {et~al.}(2021)\citenamefont {Zhao}, \citenamefont {Mintert}, \citenamefont {Moessner},\ and\ \citenamefont {Knolle}}]{Zhao2021multipolar}%
  \BibitemOpen
  \bibfield  {author} {\bibinfo {author} {\bibfnamefont {H.}~\bibnamefont {Zhao}}, \bibinfo {author} {\bibfnamefont {F.}~\bibnamefont {Mintert}}, \bibinfo {author} {\bibfnamefont {R.}~\bibnamefont {Moessner}},\ and\ \bibinfo {author} {\bibfnamefont {J.}~\bibnamefont {Knolle}},\ }\bibfield  {title} {\bibinfo {title} {{Random Multipolar Driving: Tunably Slow Heating through Spectral Engineering}},\ }\href {https://doi.org/10.1103/PhysRevLett.126.040601} {\bibfield  {journal} {\bibinfo  {journal} {Phys. Rev. Lett.}\ }\textbf {\bibinfo {volume} {126}},\ \bibinfo {pages} {040601} (\bibinfo {year} {2021})}\BibitemShut {NoStop}%
\bibitem [{\citenamefont {Zhao}\ \emph {et~al.}(2023)\citenamefont {Zhao}, \citenamefont {Knolle},\ and\ \citenamefont {Moessner}}]{Zhao2023}%
  \BibitemOpen
  \bibfield  {author} {\bibinfo {author} {\bibfnamefont {H.}~\bibnamefont {Zhao}}, \bibinfo {author} {\bibfnamefont {J.}~\bibnamefont {Knolle}},\ and\ \bibinfo {author} {\bibfnamefont {R.}~\bibnamefont {Moessner}},\ }\bibfield  {title} {\bibinfo {title} {{Temporal disorder in spatiotemporal order}},\ }\href {https://doi.org/10.1103/PhysRevB.108.L100203} {\bibfield  {journal} {\bibinfo  {journal} {Phys. Rev. B}\ }\textbf {\bibinfo {volume} {108}},\ \bibinfo {pages} {L100203} (\bibinfo {year} {2023})}\BibitemShut {NoStop}%
\bibitem [{\citenamefont {Liu}\ \emph {et~al.}(2026)\citenamefont {Liu}, \citenamefont {Liu}, \citenamefont {Liang}, \citenamefont {Deng}, \citenamefont {Chen}, \citenamefont {Shi}, \citenamefont {Li}, \citenamefont {Zhang}, \citenamefont {Chen}, \citenamefont {Fang}, \citenamefont {Feng}, \citenamefont {Gu}, \citenamefont {He}, \citenamefont {Huang}, \citenamefont {Li}, \citenamefont {Liu}, \citenamefont {Li}, \citenamefont {Mei}, \citenamefont {Peng}, \citenamefont {Song}, \citenamefont {Wang}, \citenamefont {Wang}, \citenamefont {Wang}, \citenamefont {Xiao}, \citenamefont {Xu}, \citenamefont {Xu}, \citenamefont {Yan}, \citenamefont {Yu}, \citenamefont {Yuan}, \citenamefont {Zhang}, \citenamefont {Zhao}, \citenamefont {Zhao}, \citenamefont {Zhou}, \citenamefont {Wang}, \citenamefont {Song}, \citenamefont {Tian}, \citenamefont {Mintert}, \citenamefont {Knolle}, \citenamefont {Moessner}, \citenamefont {Zhang}, \citenamefont {Zhang}, \citenamefont {Xiang}, \citenamefont {Zheng}, \citenamefont {Xu}, \citenamefont {Zhao},\ and\ \citenamefont {Fan}}]{Liu2026prethermalization}%
  \BibitemOpen
  \bibfield  {author} {\bibinfo {author} {\bibfnamefont {Z.-H.}\ \bibnamefont {Liu}}, \bibinfo {author} {\bibfnamefont {Y.}~\bibnamefont {Liu}}, \bibinfo {author} {\bibfnamefont {G.-H.}\ \bibnamefont {Liang}}, \bibinfo {author} {\bibfnamefont {C.-L.}\ \bibnamefont {Deng}}, \bibinfo {author} {\bibfnamefont {K.}~\bibnamefont {Chen}}, \bibinfo {author} {\bibfnamefont {Y.-H.}\ \bibnamefont {Shi}}, \bibinfo {author} {\bibfnamefont {T.-M.}\ \bibnamefont {Li}}, \bibinfo {author} {\bibfnamefont {L.}~\bibnamefont {Zhang}}, \bibinfo {author} {\bibfnamefont {B.-J.}\ \bibnamefont {Chen}}, \bibinfo {author} {\bibfnamefont {C.-P.}\ \bibnamefont {Fang}}, \bibinfo {author} {\bibfnamefont {D.}~\bibnamefont {Feng}}, \bibinfo {author} {\bibfnamefont {X.-Y.}\ \bibnamefont {Gu}}, \bibinfo {author} {\bibfnamefont {Y.}~\bibnamefont {He}}, \bibinfo {author} {\bibfnamefont {K.}~\bibnamefont {Huang}}, \bibinfo {author} {\bibfnamefont {H.}~\bibnamefont {Li}}, \bibinfo {author} {\bibfnamefont {H.-T.}\ \bibnamefont {Liu}}, \bibinfo {author} {\bibfnamefont {L.}~\bibnamefont {Li}}, \bibinfo {author} {\bibfnamefont {Z.-Y.}\ \bibnamefont {Mei}}, \bibinfo {author} {\bibfnamefont {Z.-Y.}\ \bibnamefont {Peng}}, \bibinfo {author} {\bibfnamefont {J.-C.}\ \bibnamefont {Song}}, \bibinfo {author} {\bibfnamefont {M.-C.}\ \bibnamefont {Wang}}, \bibinfo {author} {\bibfnamefont {S.-L.}\ \bibnamefont {Wang}}, \bibinfo {author} {\bibfnamefont {Z.}~\bibnamefont {Wang}}, \bibinfo {author} {\bibfnamefont {Y.}~\bibnamefont {Xiao}}, \bibinfo {author} {\bibfnamefont {M.}~\bibnamefont {Xu}}, \bibinfo {author} {\bibfnamefont {Y.-S.}\ \bibnamefont {Xu}}, \bibinfo {author} {\bibfnamefont {Y.}~\bibnamefont {Yan}}, \bibinfo {author} {\bibfnamefont {Y.-H.}\ \bibnamefont {Yu}}, \bibinfo {author} {\bibfnamefont {W.-P.}\ \bibnamefont {Yuan}}, \bibinfo {author} {\bibfnamefont {J.-C.}\ \bibnamefont {Zhang}}, \bibinfo {author} {\bibfnamefont {J.-J.}\ \bibnamefont {Zhao}}, \bibinfo {author} {\bibfnamefont {K.}~\bibnamefont {Zhao}}, \bibinfo {author} {\bibfnamefont {S.-Y.}\ \bibnamefont {Zhou}}, \bibinfo {author} {\bibfnamefont {Z.-A.}\ \bibnamefont {Wang}}, \bibinfo {author} {\bibfnamefont {X.}~\bibnamefont {Song}}, \bibinfo {author} {\bibfnamefont {Y.}~\bibnamefont {Tian}}, \bibinfo {author} {\bibfnamefont {F.}~\bibnamefont {Mintert}}, \bibinfo {author} {\bibfnamefont {J.}~\bibnamefont {Knolle}}, \bibinfo {author} {\bibfnamefont {R.}~\bibnamefont {Moessner}}, \bibinfo {author} {\bibfnamefont {Y.-R.}\ \bibnamefont {Zhang}}, \bibinfo {author} {\bibfnamefont {P.}~\bibnamefont {Zhang}}, \bibinfo {author} {\bibfnamefont {Z.}~\bibnamefont {Xiang}}, \bibinfo {author} {\bibfnamefont {D.}~\bibnamefont {Zheng}}, \bibinfo {author} {\bibfnamefont {K.}~\bibnamefont {Xu}}, \bibinfo {author} {\bibfnamefont {H.}~\bibnamefont {Zhao}},\ and\ \bibinfo {author} {\bibfnamefont {H.}~\bibnamefont {Fan}},\ }\bibfield  {title} {\bibinfo {title} {{Prethermalization by random multipolar driving on a 78-qubit processor}},\ }\href {https://doi.org/10.1038/s41586-025-09977-x} {\bibfield  {journal} {\bibinfo  {journal} {Nature}\ }\textbf {\bibinfo {volume} {650}},\ \bibinfo {pages} {79} (\bibinfo {year} {2026})}\BibitemShut {NoStop}%
\bibitem [{\citenamefont {Hastings}\ and\ \citenamefont {Haah}(2021)}]{Haah21honeycomb}%
  \BibitemOpen
  \bibfield  {author} {\bibinfo {author} {\bibfnamefont {M.~B.}\ \bibnamefont {Hastings}}\ and\ \bibinfo {author} {\bibfnamefont {J.}~\bibnamefont {Haah}},\ }\bibfield  {title} {\bibinfo {title} {{Dynamically {G}enerated {L}ogical {Q}ubits}},\ }\href {https://doi.org/10.22331/q-2021-10-19-564} {\bibfield  {journal} {\bibinfo  {journal} {{Quantum}}\ }\textbf {\bibinfo {volume} {5}},\ \bibinfo {pages} {564} (\bibinfo {year} {2021})}\BibitemShut {NoStop}%
\bibitem [{\citenamefont {Aasen}\ \emph {et~al.}(2022)\citenamefont {Aasen}, \citenamefont {Wang},\ and\ \citenamefont {Hastings}}]{Hastings22honeycomb}%
  \BibitemOpen
  \bibfield  {author} {\bibinfo {author} {\bibfnamefont {D.}~\bibnamefont {Aasen}}, \bibinfo {author} {\bibfnamefont {Z.}~\bibnamefont {Wang}},\ and\ \bibinfo {author} {\bibfnamefont {M.~B.}\ \bibnamefont {Hastings}},\ }\bibfield  {title} {\bibinfo {title} {{Adiabatic paths of Hamiltonians, symmetries of topological order, and automorphism codes}},\ }\href {https://doi.org/10.1103/PhysRevB.106.085122} {\bibfield  {journal} {\bibinfo  {journal} {Phys. Rev. B}\ }\textbf {\bibinfo {volume} {106}},\ \bibinfo {pages} {085122} (\bibinfo {year} {2022})}\BibitemShut {NoStop}%
\bibitem [{\citenamefont {Davydova}\ \emph {et~al.}(2023)\citenamefont {Davydova}, \citenamefont {Tantivasadakarn},\ and\ \citenamefont {Balasubramanian}}]{Nat23floquetwithoutparent}%
  \BibitemOpen
  \bibfield  {author} {\bibinfo {author} {\bibfnamefont {M.}~\bibnamefont {Davydova}}, \bibinfo {author} {\bibfnamefont {N.}~\bibnamefont {Tantivasadakarn}},\ and\ \bibinfo {author} {\bibfnamefont {S.}~\bibnamefont {Balasubramanian}},\ }\bibfield  {title} {\bibinfo {title} {{Floquet Codes without Parent Subsystem Codes}},\ }\href {https://doi.org/10.1103/PRXQuantum.4.020341} {\bibfield  {journal} {\bibinfo  {journal} {PRX Quantum}\ }\textbf {\bibinfo {volume} {4}},\ \bibinfo {pages} {020341} (\bibinfo {year} {2023})}\BibitemShut {NoStop}%
\bibitem [{\citenamefont {{Zhu}}\ and\ \citenamefont {{Trebst}}(2023)}]{Zhu23qubit}%
  \BibitemOpen
  \bibfield  {author} {\bibinfo {author} {\bibfnamefont {G.-Y.}\ \bibnamefont {{Zhu}}}\ and\ \bibinfo {author} {\bibfnamefont {S.}~\bibnamefont {{Trebst}}},\ }\bibfield  {title} {\bibinfo {title} {{Qubit fractionalization and emergent Majorana liquid in the honeycomb Floquet code induced by coherent errors and weak measurements}},\ }\href@noop {} {\bibfield  {journal} {\bibinfo  {journal} {preprint}\ } (\bibinfo {year} {2023})},\ \Eprint {https://arxiv.org/abs/2311.08450} {arXiv:2311.08450} \BibitemShut {NoStop}%
\bibitem [{\citenamefont {Eckstein}\ \emph {et~al.}(2026)\citenamefont {Eckstein}, \citenamefont {Schmid}, \citenamefont {Preiss}, \citenamefont {Trebst}, \citenamefont {von Oppen},\ and\ \citenamefont {Zhu}}]{zenodo_fibonacci}%
  \BibitemOpen
  \bibfield  {author} {\bibinfo {author} {\bibfnamefont {F.}~\bibnamefont {Eckstein}}, \bibinfo {author} {\bibfnamefont {H.}~\bibnamefont {Schmid}}, \bibinfo {author} {\bibfnamefont {Q.}~\bibnamefont {Preiss}}, \bibinfo {author} {\bibfnamefont {S.}~\bibnamefont {Trebst}}, \bibinfo {author} {\bibfnamefont {F.}~\bibnamefont {von Oppen}},\ and\ \bibinfo {author} {\bibfnamefont {G.-Y.}\ \bibnamefont {Zhu}},\ }\bibfield  {title} {\bibinfo {title} {{Data for ``Dynamical self-dual criticality in Fibonacci-monitored quantum Ising chains"}},\ }\bibfield  {journal} {\bibinfo  {journal} {Zenodo}\ }\href {https://doi.org/10.5281/zenodo.20266376} {10.5281/zenodo.20266376} (\bibinfo {year} {2026})\BibitemShut {NoStop}%
\bibitem [{\citenamefont {Magnus}(1954)}]{Magnus1954}%
  \BibitemOpen
  \bibfield  {author} {\bibinfo {author} {\bibfnamefont {W.}~\bibnamefont {Magnus}},\ }\bibfield  {title} {\bibinfo {title} {{On the exponential solution of differential equations for a linear operator}},\ }\href {https://doi.org/10.1002/cpa.3160070404} {\bibfield  {journal} {\bibinfo  {journal} {Commun. Pure Appl. Math.}\ }\textbf {\bibinfo {volume} {7}},\ \bibinfo {pages} {649} (\bibinfo {year} {1954})}\BibitemShut {NoStop}%
\bibitem [{\citenamefont {Blanes}\ \emph {et~al.}(2009)\citenamefont {Blanes}, \citenamefont {Casas}, \citenamefont {Oteo},\ and\ \citenamefont {Ros}}]{Blanes2009magnus}%
  \BibitemOpen
  \bibfield  {author} {\bibinfo {author} {\bibfnamefont {S.}~\bibnamefont {Blanes}}, \bibinfo {author} {\bibfnamefont {F.}~\bibnamefont {Casas}}, \bibinfo {author} {\bibfnamefont {J.~A.}\ \bibnamefont {Oteo}},\ and\ \bibinfo {author} {\bibfnamefont {J.}~\bibnamefont {Ros}},\ }\bibfield  {title} {\bibinfo {title} {{The Magnus expansion and some of its applications}},\ }\href {https://doi.org/10.1016/j.physrep.2008.11.001} {\bibfield  {journal} {\bibinfo  {journal} {Phys. Rep.}\ }\textbf {\bibinfo {volume} {470}},\ \bibinfo {pages} {151} (\bibinfo {year} {2009})}\BibitemShut {NoStop}%
\bibitem [{\citenamefont {Sykes}\ and\ \citenamefont {Essam}(1963)}]{Sykes1963}%
  \BibitemOpen
  \bibfield  {author} {\bibinfo {author} {\bibfnamefont {M.~F.}\ \bibnamefont {Sykes}}\ and\ \bibinfo {author} {\bibfnamefont {J.~W.}\ \bibnamefont {Essam}},\ }\bibfield  {title} {\bibinfo {title} {{Some Exact Critical Percolation Probabilities for Bond and Site Problems in Two Dimensions}},\ }\href {https://doi.org/10.1103/PhysRevLett.10.3} {\bibfield  {journal} {\bibinfo  {journal} {Phys. Rev. Lett.}\ }\textbf {\bibinfo {volume} {10}},\ \bibinfo {pages} {3} (\bibinfo {year} {1963})}\BibitemShut {NoStop}%
\bibitem [{\citenamefont {Wan}\ \emph {et~al.}(2026)\citenamefont {Wan}, \citenamefont {Dai},\ and\ \citenamefont {Zhu}}]{Wan25nishimori}%
  \BibitemOpen
  \bibfield  {author} {\bibinfo {author} {\bibfnamefont {Z.-Q.}\ \bibnamefont {Wan}}, \bibinfo {author} {\bibfnamefont {X.-D.}\ \bibnamefont {Dai}},\ and\ \bibinfo {author} {\bibfnamefont {G.-Y.}\ \bibnamefont {Zhu}},\ }\bibfield  {title} {\bibinfo {title} {{Revisiting Nishimori multicriticality through the lens of information measures}},\ }\href {https://doi.org/10.1103/b8y5-k3y6} {\bibfield  {journal} {\bibinfo  {journal} {Phys. Rev. Res.}\ }\textbf {\bibinfo {volume} {8}},\ \bibinfo {pages} {023059} (\bibinfo {year} {2026})}\BibitemShut {NoStop}%
\end{thebibliography}%
%%%%%%%%%%%%%%%%%%%%%%%%%%%%%%%%%%%%%%%%%%%%%%%%%%%%%%%%%%%%%%%%%%%

%%%%%%%%%%%%%%%%%%%%%%%%%%%%%%%%%%%%%%%%%%%%%%%%%%%%%%%%%%%%%%%%%%%
%  Appendix
%%%%%%%%%%%%%%%%%%%%%%%%%%%%%%%%%%%%%%%%%%%%%%%%%%%%%%%%%%%%%%%%%%%

\clearpage
\appendix

%%%%%%%%%%%%%%%%%%%%%%%%%%%%%%%%%%%%%%%%%%%%%%%%%%%%%%%%%%%%%%%%%%%
% Supplemental data
%%%%%%%%%%%%%%%%%%%%%%%%%%%%%%%%%%%%%%%%%%%%%%%%%%%%%%%%%%%%%%%%%%%
\section{Supplemental data}

%%%%%%%%%%%%%%%%%%%%%%%%%%%%%%%%%%%%%%%%%%%%%%%%%%%%%%%%%%%%%%%%%%%
\subsection*{Phase diagram numerics and critical behavior}
%%%%%%%%%%%%%%%%%%%%%%%%%%%%%%%%%%%%%%%%%%%%%%%%%%%%%%%%%%%%%%%%%%%

%%%%%%%%%%%%%%%%%%%%%%%%%%%%%%%%%%%%%%%%%%%%%%%%%%%%%%%%%%%%%%%%%%%
\subsubsection*{Coherent information and critical length exponent}
%%%%%%%%%%%%%%%%%%%%%%%%%%%%%%%%%%%%%%%%%%%%%%%%%%%%%%%%%%%%%%%%%%%
The phase diagram in Fig.~\ref{fig:QuantumCircuit}(c) can be diagnosed by the coherent information of the one-dimensional chain, with a LRE state such as the GHZ state serving as a logical memory. The coherent information then indicates, whether this logical memory is still intact at the end of the circuit, or whether it (or a part of it) has been lost. Viewing the GHZ state as a repetition code, the coherent information is determined by entangling a reference qubit R with the initial state state of the qubit chain to form a joint GHZ state $|\psi_0 \rangle _{QR} = |00...00\rangle_Q \otimes |0\rangle_R + |11...11\rangle_Q \otimes |1\rangle_R$. The coherent information is then given by 
\begin{equation}
    I_c = S_{vN}(\rho_Q) - S_{vN}(\rho_{QR})
\end{equation}
It indicates the amount of information retained in the chain. 
Our  setup here is similar to the one of Ref.~\cite{Wang25selfdual}, but we significantly increase the system sizes in order to pinpoint the critical location 
for {\it emergent} self-duality not present on the microscopic level.

\begin{figure}[t!]
    \centering
    \includegraphics[width=0.95\columnwidth]{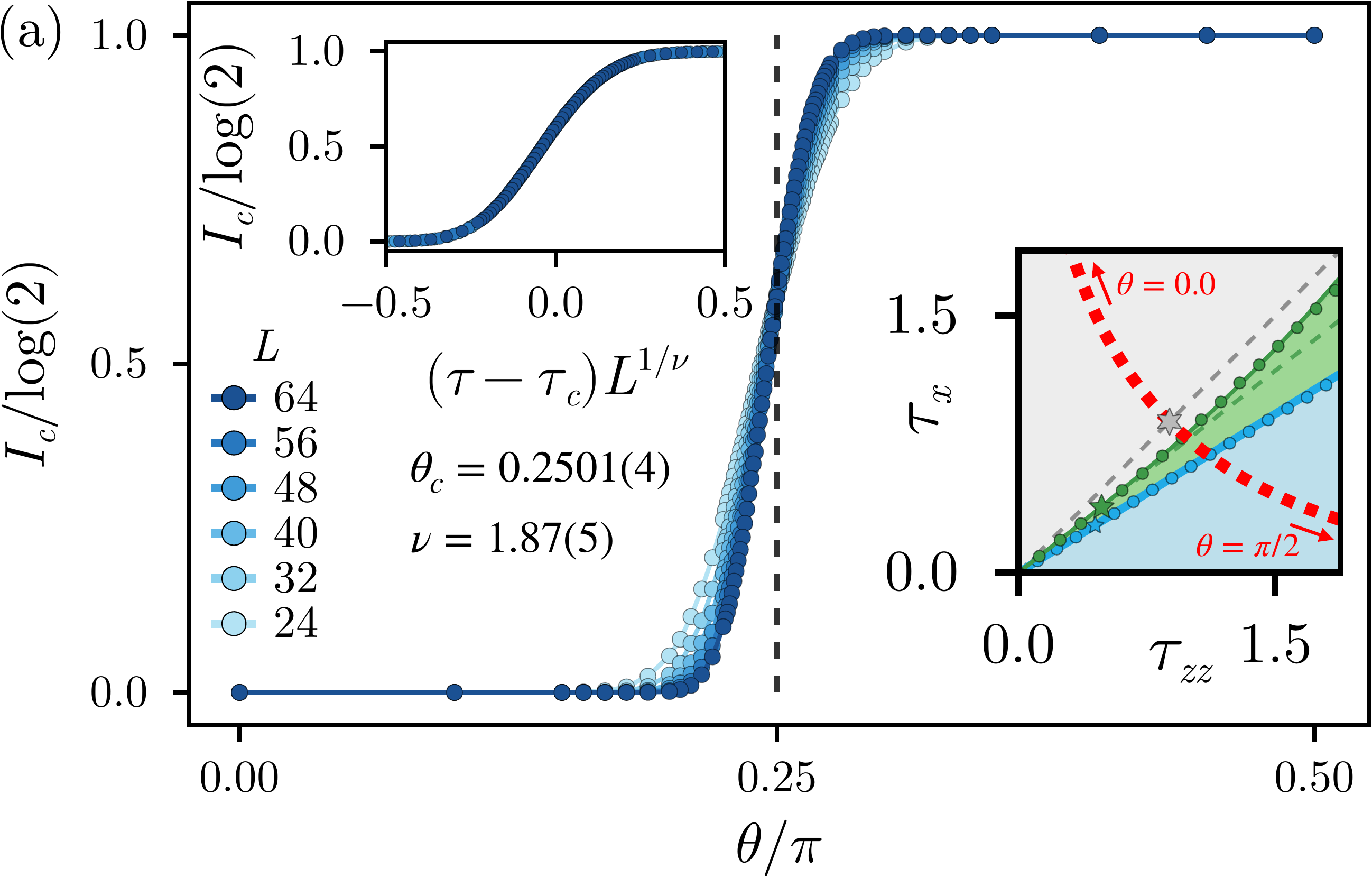}
    \includegraphics[width=0.95\columnwidth]{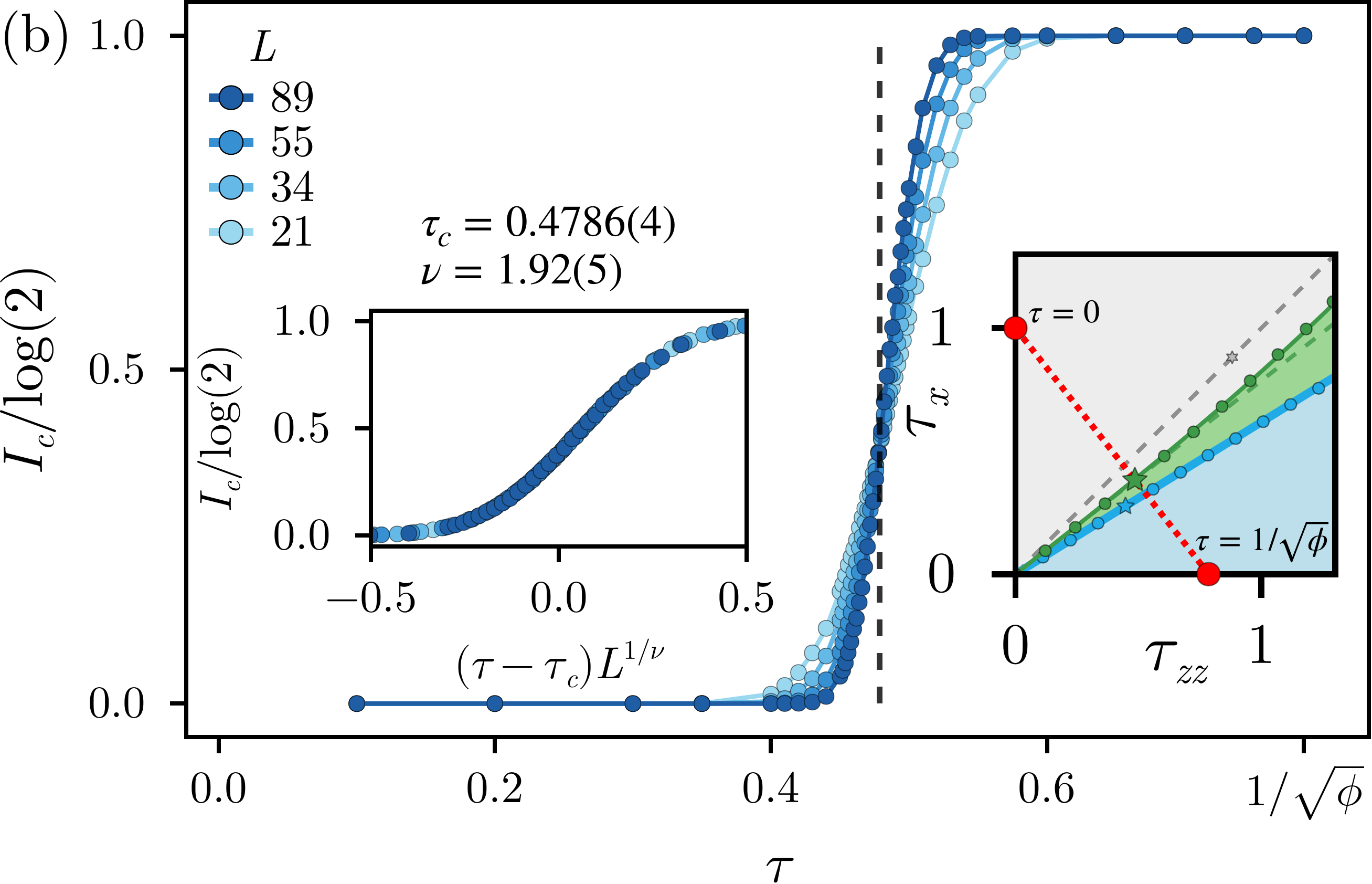}
    \includegraphics[width=0.95\columnwidth]{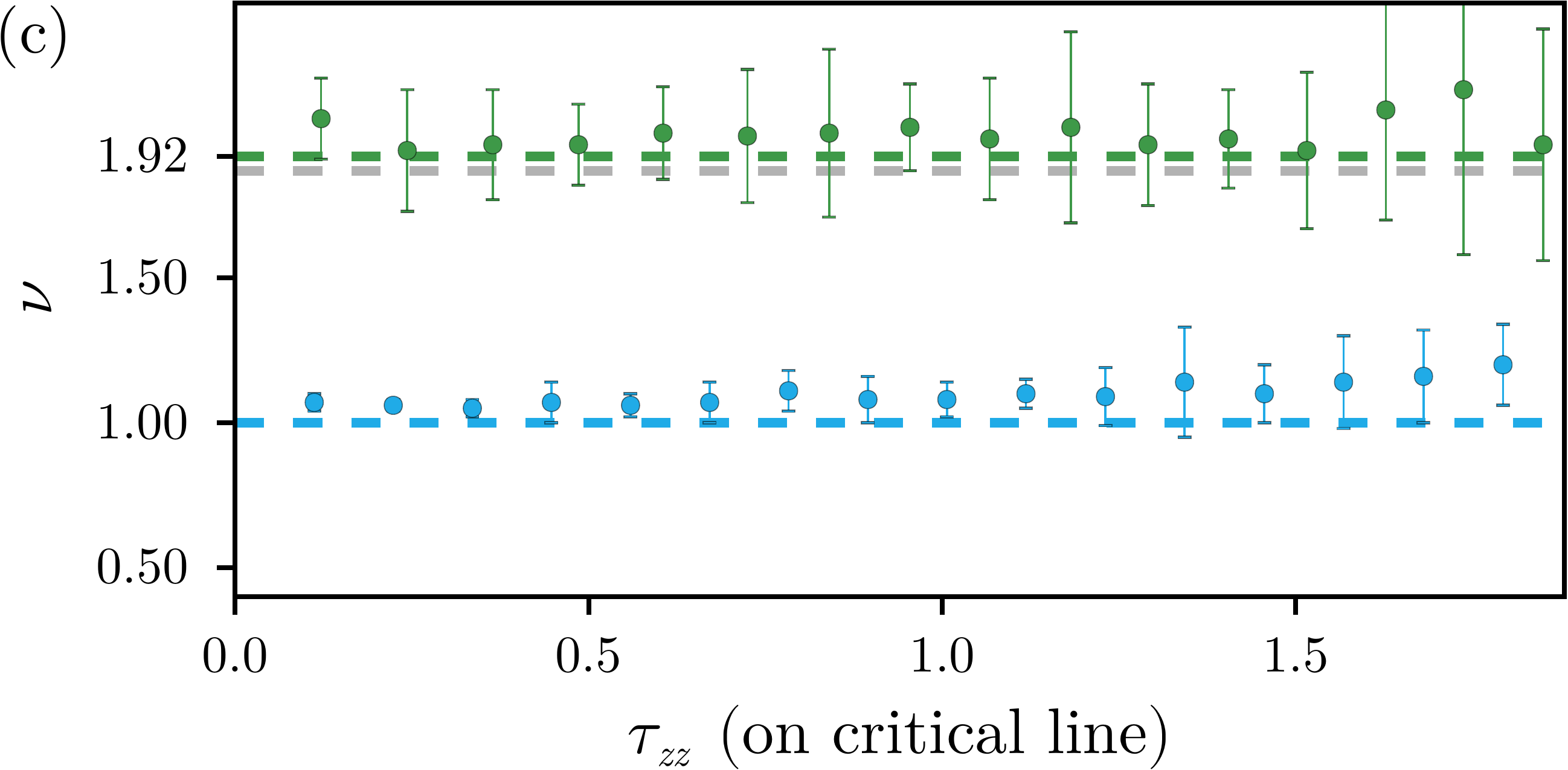}
    \caption{{\bf Critical length exponent of the coherent information.} 
    (a) Phase transition and finite-size scaling of the coherent information for the Floquet circuit. The parametrization of $\theta$ realizes the space-time dual of the Floquet circuit (marked in red on the right). The finite-size scaling obtains a critical length exponent of $\nu=1.87(5)$. (b) Phase transition and finite-size scaling of the coherent information for the Fibonacci circuit. $\tau$ follows a linear sweep through the phase diagram (marked in red on the right). The finite-size scaling obtains $\nu=1.92(5)$. The two critical length exponents thus agree with each other supporting that both transitions share the same universality class. (c)  The critical exponent of multiple finite-size scaling on the Born averaged critical line (green) and the post-selected critical line (blue). The $\nu$ on the Born averaged critical line agrees with the more precise finite-size scalings of the Floquet ((a), grey) and the Fibonacci circuit ((b), green). On the post-selected critical line $\nu$ agrees with the Ising exponent $\nu=1$. On both critical lines the critical exponents stay approximately constant, indicating that the complete line shares the same universality class.}
    \label{fig:Endmatter_comparison_nu}
\end{figure}

Projective measurement of the chain would collapse the 
LRE and reduce the coherent information to zero. An observer could then completely construct the logical information out of the measurement outcomes, while there is no information left in the state of the chain itself. In contrast, for vanishing measurement strength the measurement outcomes do not obtain any information. Instead the 
LRE phase is preserved and the logical memory still stored in the chain, thus obtaining a coherent information of $I_c=\log(2)$. 
$M_{zz}$ measurements form LRE and nudge the chain towards a (glassy) GHZ state, while $M_x$ measurements reduce the entanglement and nudge the chain towards a trivial product state. Therefore the coherent information tends to one in the blue (post-selected) or blue and green (Born average) region, while it tends to zero in the gray (Born average) or gray and green (post-selected) region. The transition in between can be directly observed in the coherent information as a crossing point of different system sizes with constant time-space ratio. Figure \ref{fig:Endmatter_comparison_nu} shows this transition for the Floquet circuit (a) and the Fibonacci circuit (b) as a direct comparison. The crossing point can be determined by a finite-size scaling analysis, where the parameter is rescaled by $(\theta-\theta_c)L^{1/\nu}$ (Floquet) or $(\tau-\tau_c)L^{1/\nu}$ (Fibonacci). In particular, this also accesses the scaling dimension of the coherent information, the critical length exponent $\nu$. The obtained values $\nu=1.87(5)$ for the Floquet circuit and $\nu=1.92(5)$ for the Fibonacci circuit are in reasonable agreement with one another, which supports that both transitions are of the same universality class. Furthermore, Fig.~\ref{fig:Endmatter_comparison_nu} (c) shows the critical length exponent for different measurement strengths along the post-selected (blue) and Born average (green) critical line. On both critical lines $\nu$ stays constant up to numerical fluctuations. On the post-selected critical line, it  agrees with the value $\nu=1$ for Ising transitions, on the Born average critical line it agrees with the more precise finite size scalings of (a) and (b). The numerical estimates for the exponent $\nu$ obtained here improve the previous estimate  $\nu=1.72(8)$ of Ref.~\cite{Wang25selfdual} 
obtained for the Floquet circuit for modest system sizes $L = 4,8,16,32$.

%%%%%%%%%%%%%%%%%%%%%%%%%%%%%%%%%%%%%%%%%%%%%%%%%%%%%%%%%%%%%%%%%%%
\subsubsection*{Entanglement scaling on critical line of Born weak measurement}
%%%%%%%%%%%%%%%%%%%%%%%%%%%%%%%%%%%%%%%%%%%%%%%%%%%%%%%%%%%%%%%%%%%

%%%%%%%%%%%%%%%%%%%%%%%%%%%%%%%%%%%%%%%%%%%%%%%%%%%%%%%%%%%%%%%%%%%
\begin{figure}[b!]
    \centering
    \includegraphics[width=\columnwidth]{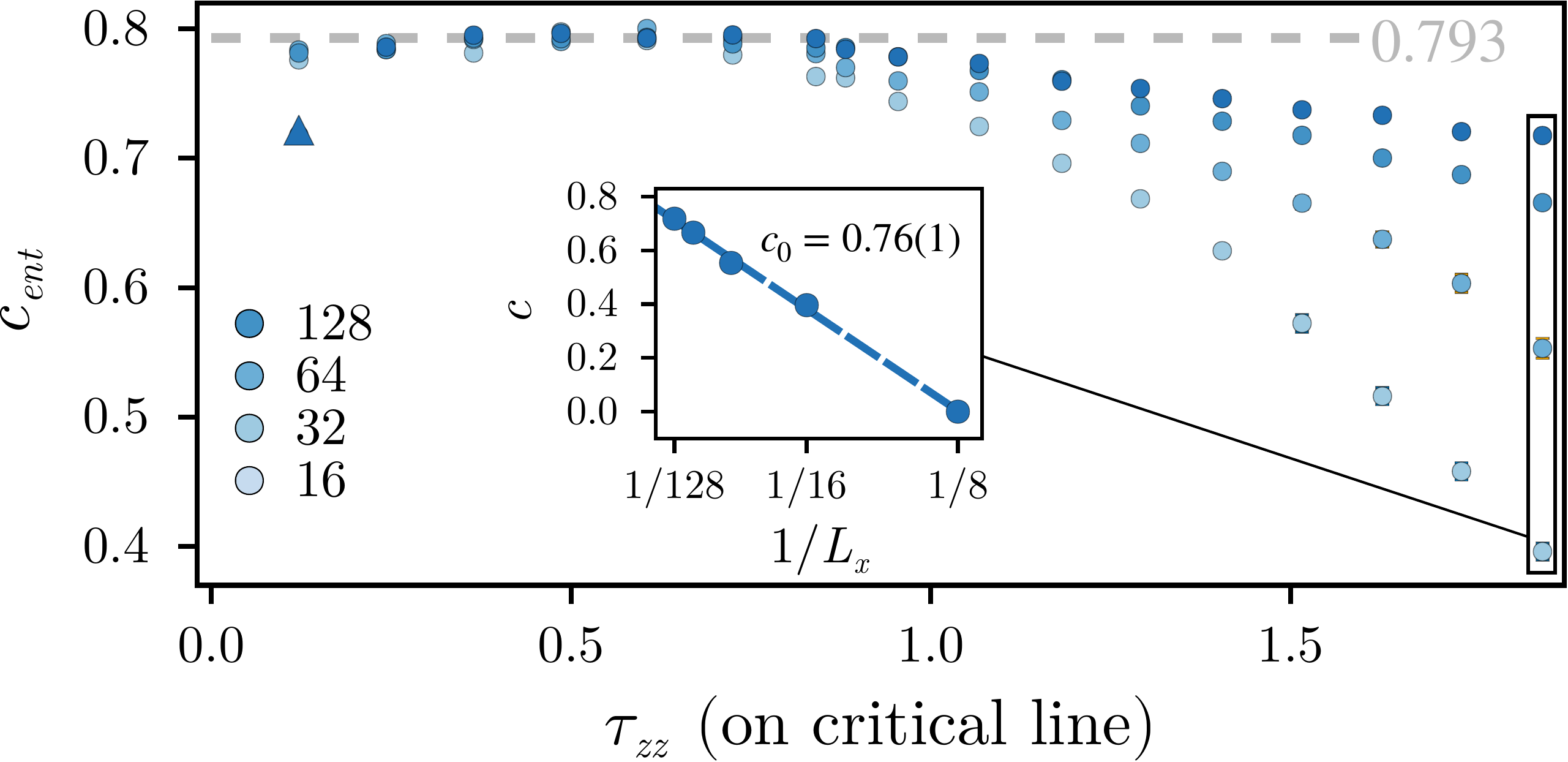}
    \caption{ {\bf Entanglement scaling dimension $c_{\rm ent}$ on the Born average critical line.} 
    The entanglement scaling dimension extracted from fits of the entanglement arc for each system according to Eq.~\ref{eq:entanglementarcs}. 
    $c_{\rm ent}$ tends to $0.793(1)$ as observed in the precise scaling of Fig.~\ref{fig:EntropyScaling_combined}. 
    For strong measurements we observe that the finite-size effects grow larger. 
    $L=128$ thus does not observe the scaling dimension of the thermodynamic limit. 
    The triangle point for weakest measurement strength but largest size $L=128$ has not reached steady state even after 4181 layers of gates. 
    }
    \label{fig:Centralcharge_on_criticalline}
\end{figure}
%%%%%%%%%%%%%%%%%%%%%%%%%%%%%%%%%%%%%%%%%%%%%%%%%%%%%%%%%%%%%%%%%%%

The scaling of the entanglement arc of a system on one of the critical lines can be described by the conformal field theory (CFT) prediction~\cite{Calabrese2004,calabrese2009entanglement}
\begin{equation} \label{eq:entanglementarcs}
    S_{\mathrm{CFT}}(l) = \frac{c_{\rm ent}}{3}\ln\!\left[\frac{L}{\pi}\sin\!\left(\frac{\pi l}{L}\right)\right]
\end{equation}
for periodic boundary conditions at criticality. For a unitary CFT, $c_{\rm ent}$ is the central charge, while for measurement-induced criticality it captures the scaling dimension of the boundary-condition-changing operator~\cite{Ludwig2020}. 

\begin{figure}[t!]
    \centering
    \includegraphics[width=\columnwidth]{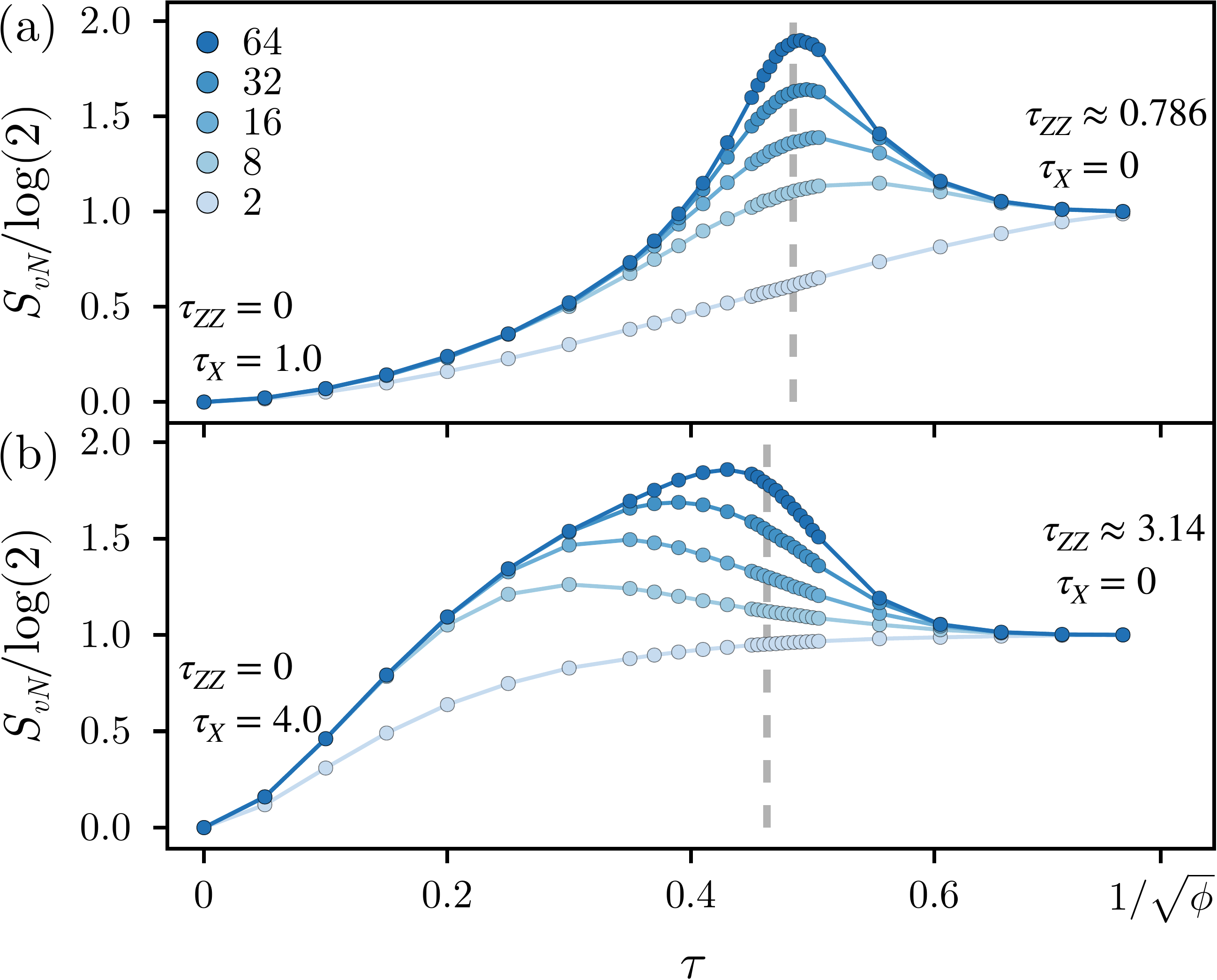}
    \caption{ {\bf Born average entanglement entropy transition.} 
    (a) Born average entanglement entropy of a weak measurement sweep going from $\tau_{zz}=0$ and $\tau_x=1.0$ to $\tau_{zz}\approx0.786$.
    (b) Born average entanglement entropy of sweep of stronger weak measurement going from $\tau_{zz}=0$ and $\tau_x=4.0$ to $\tau_{zz}\approx3.14$.
    The behavior of the entanglement entropy changes around the transition with larger measurement strength. 
    In addition to a small shift of the transition matching the critical line, the finite-size effects increase, with the entanglement entropy peak of finite systems showing larger deviations from the transition for increasing measurement strength. Nonetheless, in the thermodynamic limit the peak always tends to the transition.
    }
    \label{fig:Peakcomparison}
\end{figure}

By fitting the numerical data for different measurement strengths on the Born average critical line, we observe the behavior of $c_{\rm ent}$. We find that $c_{\rm ent}$ generally approaches the result of $c_{\rm ent}=0.793(1)$ of Fig.~\ref{fig:EntropyScaling_combined}. For larger measurement strengths, $c_{\rm ent}$ exhibits strong finite-size effects. The largest simulated system size, $L=128$, still shows significant deviation from 0.793. The central charge has not yet saturated and continues to shift with increasing system size. Extrapolating the shift of $c_{\rm ent}$ yields $c_{\rm ent,0}=0.76(1)$, which approximately recovers a similar scaling dimension. Up to these finite-size effects, $c_{\rm ent}$ stays approximately constant, signaling that there is no change in universality class. The nature of the finite-size effects is displayed in Fig.~\ref{fig:Peakcomparison}: For the measurement strength with precisely $c_{\rm ent}=0.793(1)$, the peak of the entanglement entropy is close to the transition already for system sizes of $L=16$ to $L=64$ [Fig.~\ref{fig:Peakcomparison}(a)]. In contrast, for larger measurement strengths the peak has a finite-size shift [Fig.~\ref{fig:Peakcomparison}(b)]. Extrapolation shows that the peak position of $S_{vN}$ still matches the transition location in the thermodynamic limit, but finite system sizes experience a deviation that leads to a changed value of $c_{\rm ent}$. It should also be noted that the number of time steps needs to be sufficient for the system to reach equilibrium. The simulations in this paper utilize $L_y=4181$ (layers of gates in the circuit), which is enough for equilibrium to be reached in all of the data points, except the very first data point for very weak measurement strength and our largest system size $L=128$ marked as a triangle, which signals that lower measurement strengths would require a larger number of time steps to reach equilibrium.

%%%%%%%%%%%%%%%%%%%%%%%%%%%%%%%%%%%%%%%%%%%%%%%%%%%%%%%%%%%%%%%%%%%
%%%%%%%%%%%%%%%%%%%%%%%%%%%%%%%%%%%%%%%%%%%%%%%%%%%%%%%%%%%%%%%%%%%

%%%%%%%%%%%%%%%%%%%%%%%%%%%%%%%%%%%%%%%%%%%%%%%%%%%%%%%%%%%%%%%%%%%
\subsubsection*{Fourier peak scaling exponent along the critical line}
%%%%%%%%%%%%%%%%%%%%%%%%%%%%%%%%%%%%%%%%%%%%%%%%%%%%%%%%%%%%%%%%%%%

\begin{figure}[t!]
    \centering
    \includegraphics[width=\columnwidth]{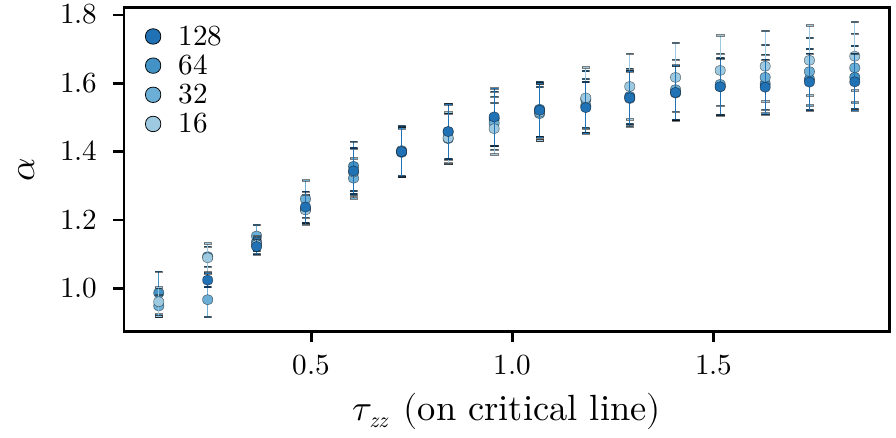}
    \caption{ {\bf Fourier peak scaling exponent $\alpha$ along the critical line of Born weak measurements.} 
    The exponent $\alpha$ is taken from linear fits to the Fibonacci-numbered peaks of the Fourier transform of the entanglement entropy as in Fig.~3 in the main text.
    }
    \label{fig:AlphaOnCriticalline}
\end{figure}

We consider the Fourier spectrum of the time evolution of the entanglement entropy and determine the scaling behavior of the Fourier peaks as in Fig.~\ref{fig:EntanglementDynamics}. As detailed there, the amplitudes of the Fourier peaks decay as a power law of the form $ |F_{\omega_n} [S_{vN}]|\propto   n^{-\alpha}$ for integer multiples of the golden ratio $\omega_n=\frac{2\pi n}{\phi}$. In the projective limit, where the entanglement entropy exactly  tracks the Fibonacci word, the scaling exponent evaluates to $\alpha=2$ for Fibonacci numbers and oscillating behavior bounded by $\alpha=1$ for non-Fibonacci numbers. Fitting the power law of the Fibonacci-numbered peaks determines characteristic values of $\alpha$ for finite measurement strengths. This indicates that $\alpha$ exhibits a monotonic and continuous increase (Fig.~\ref{fig:AlphaOnCriticalline}).

%%%%%%%%%%%%%%%%%%%%%%%%%%%%%%%%%%%%%%%%%%%%%%%%%%%%%%%%%%%%%%%%%%%
%%%%%%%%%%%%%%%%%%%%%%%%%%%%%%%%%%%%%%%%%%%%%%%%%%%%%%%%%%%%%%%%%%%

%%%%%%%%%%%%%%%%%%%%%%%%%%%%%%%%%%%%%%%%%%%%%%%%%%%%%%%%%%%%%%%%%%%
 \subsection*{Temporal correlation of measurement outcomes}
 %%%%%%%%%%%%%%%%%%%%%%%%%%%%%%%%%%%%%%%%%%%%%%%%%%%%%%%%%%%%%%%%%%%

\begin{figure}[t!]
     \centering
     \includegraphics[width=\columnwidth]{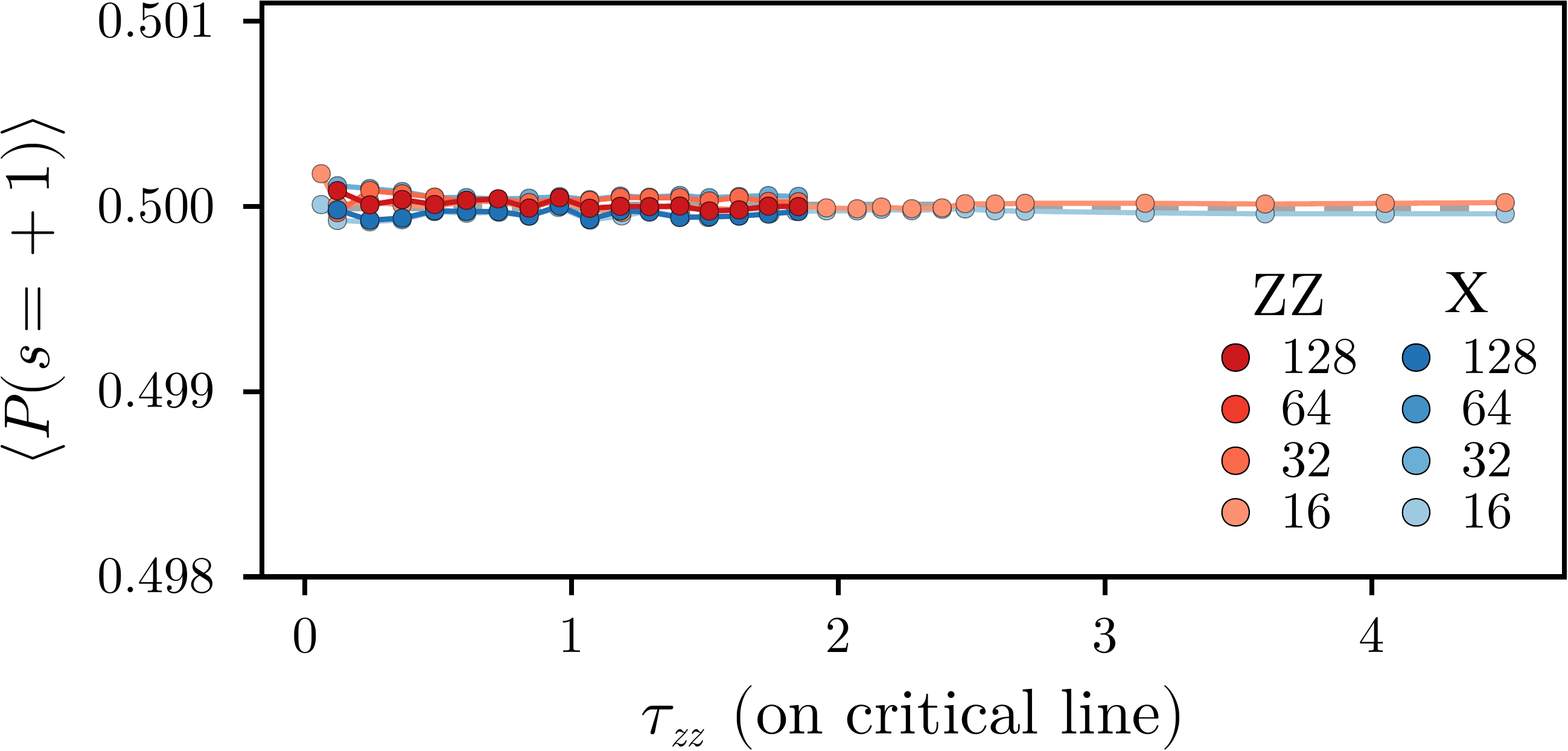}
     \caption{{\bf Probability for a measurement outcome of +1 on the critical line.} 
     $P(s=+1)$ gives the probability for a measurement outcome being +1, averaged over time and space 
     (with no particular additional features in time or space). Due to their different occurrence, $M_{zz}$ and $M_x$ have distinct behavior.
     The dashed lines signal $P=0.5$, so measurement outcomes are simply equally distributed,
     which seems to be the case for the complete critical line (with numerical fluctuation on the order of $10^{-4}$). 
    The data are based on the measurement outcomes for $L=128$, $L=64$, $L=32$ and $L=16$ on the critical line of the Born average.
     $L=16$ is expanded further to higher measurement strengths with an estimation of transition values.
     }
     \label{fig:Disorder_pureaverage}
 \end{figure}

Beyond the quasiperiodic arrangement of gates in our circuit protocol, the randomness arising from the measurement operations gives rise to an ensemble of individual trajectories. For multiple runs of the circuit, this ensemble is captured by the distribution of measurement records (along the trajectories, corresponding to the bulk of the spacetime evolution). Since the two types of measurements $M_{zz}$ and $M_x$ are distinct, we analyze their respective measurement records individually. Averaging the record over time and space shows that there are no biases in such a direct average: On average both measurement results are equally probable for any system size and measurement strength, as Fig.~\ref{fig:Disorder_pureaverage} illustrates. There the average probability of a measurement result $s=+1$ is shown for system sizes $L=16$ to $L=128$ on the Born average critical line. For system sizes $L=32$ to $L=128$, data are restricted to the exactly determined critical points. For $L=16$, additional data for estimated transition locations are provided. All data points stay close to $P(s=+1)=\frac{1}{2}$ and no particular features are exhibited in this average.

Beyond the average value, we investigate the temporal correlations within the measurement record. Figure \ref{fig:TimeCorrelationMeasurements}(a) plots the probability that subsequent measurements of the same operator give identical result.  Clear correlations between measurements in time can be observed. We observe that the two measurement types have clearly distinct behavior since two $M_x$ measurements can directly follow after one another in the Fibonacci sequence, while $M_{zz}$ measurements are always separated by a layer of $M_x$ measurements. In the projective limit, this leads to uncorrelated measurement results for $M_{zz}$, while consecutive $M_x$ measurements either give uncorrelated results, if they sandwich a layer of $M_{zz}$ measurements, or strongly correlated measurement outcomes, if there are no intervening $M_{zz}$ measurements. The ratio of such directly consecutive time steps in the Fibonacci sequence determines the projective limit of $M_x$ to be $P_{X,\infty}=1-\frac{1}{\phi}+\frac{1}{2\phi}\approx 0.69$. While $M_{zz}$ displays completely random measurement results in both limits, a peak of time correlation at finite measurement strength can be observed. The behavior is independent of system size, with all systems displaying the same numerical behavior.

%%%%%%%%%%%%%%%%%%%%%%%%%%%%%%%%%%%%%%%%%%%%%%%%%%%%%%%%%%%%%%%%%%%
%%%%%%%%%%%%%%%%%%%%%%%%%%%%%%%%%%%%%%%%%%%%%%%%%%%%%%%%%%%%%%%%%%%

%%%%%%%%%%%%%%%%%%%%%%%%%%%%%%%%%%%%%%%%%%%%%%%%%%%%%%%%%%%%%%%%%%%
\section{Fourier transform of the entanglement entropy}
\label{sec:FT drive}
%%%%%%%%%%%%%%%%%%%%%%%%%%%%%%%%%%%%%%%%%%%%%%%%%%%%%%%%%%%%%%%%%%%

We consider the dynamics of the bipartite entanglement entropy in the projective limit $\tau =\infty$ in protocol (II). It exactly tracks the Fibonacci word,
\begin{align}
    S_{vN}(t)=\left\lfloor \frac{t+1}{\phi}\right\rfloor
    -
    \left\lfloor \frac{t}{\phi}\right\rfloor,
    \qquad t\in\mathbb{N},
\end{align}
where $\phi=(1+\sqrt5)/2$ denotes the golden ratio.  Throughout this Appendix, the entropy is measured in units of $\ln 2$. We derive its Fourier transform
\begin{align}
    F[S_{vN}](\omega)
    =
\sum_{t=1}^{\infty}
    e^{i\omega t} S_{vN}(t) \,.
\end{align}

To obtain analytical expressions, we first continue to $t\in\mathbb{R}$, where $\tilde{S}_{vN}(t)=\tilde{S}_{vN}(t+\phi)$ becomes periodic with period $\phi$ and assumes the form of a square wave. Hence, the spectrum contains harmonics at $\omega_n=2\pi n/\phi$. We expand in a Fourier series,
\begin{align}
\tilde{S}_{vN}(t)=\sum_{n= -\infty}^\infty F_{\omega_n}[\tilde{S}_{vN}] e^{i\omega_n t} \,,
    \end{align}
with Fourier coefficients
$F_{\omega_n}[\tilde{S}_{vN}]=\frac{1}{\phi}\int^\phi_0 \dd{t}
\tilde{S}_{vN}(t) e^{-i\omega_n t}$. We find for the Fourier coefficients
\begin{align}
    F_{\omega_n}[\tilde{S}_{vN}]
     =
     \frac{1}{ n \pi}e^{-\frac{in \pi }{\phi^2}}
     \sin(\frac{n \pi}{\phi^2}) \,.
     \label{eq:Fourier coeff}
\end{align}

Next, we again restrict time to integer times $t$ where the drive is quasiperiodic. We write
\begin{align}
    F_{\omega_n}[S_{vN}](\omega)
    =
\sum^\infty_{n=-\infty}  F_{\omega_n}[\tilde{S}_{vN}](\omega)\sum^\infty_{t=1}e^{i(\omega +\omega_n )t} \,.
\end{align}
We use Poisson summation and obtain $\sum^\infty_{t=1}e^{i\alpha t}=\frac{1}{2}
\sum^\infty_{m=-\infty} (\Delta_m(\alpha)-1)$. The function $\Delta_m(\alpha)=i\alpha/(\alpha^2-(2\pi m)^2)$ has poles at $\alpha=2\pi m$ and acts as a delta function. Thus, restricting to integer times simply broadens the Fourier transform as
\begin{align}
F[S_{vN}]=  \frac{1}{2}\sum^\infty_{n,m=-\infty}
   F_{\omega_n}[\tilde{S}_{vN}]\bigg(
\Delta_m(\omega+\omega_n)-1\bigg) \,.
\end{align}

The factor  $1/n$ in Eq.\ \eqref{eq:Fourier coeff} is the characteristic Fourier scaling of a square pulse. However, for Fibonacci harmonics $n=f_k$ ($k \in \mathbb{N}$), one finds a different power law. Here, the argument of the oscillating sine-factor becomes anomalously small, since $f_k/\phi^2 \simeq f_{k-2}$ approaches an integer exponentially fast. To quantify the error, we use the Binet formula $f_{k}=(\phi^{k+1}-\psi^{k+1})/(\phi-\psi)$ with $\psi=-1/\phi$ and $f_0=f_1=1$. Expanding for large $k$ gives
\begin{align}
\frac{f_{k}}{\phi^2}= f_{k-2}  -\frac{(-1)^k}{\sqrt{5}f_{k}}+\order{f^{-3}_{k}} \,.
\end{align}
Thus, the sine in Eq.~\eqref{eq:Fourier coeff} scales as $1/f_k$, leading to an additional suppression of the Fourier coefficients at Fibonacci harmonics
 \begin{align}
|F_{\omega_{n=f_k}}[\tilde{S}_{vN}]|
     \simeq 
     \frac{1}{ \sqrt{5}\, \pi n^2 } \,.
 \end{align}
 Hence, while generic harmonics decay as $1/n$, Fibonacci harmonics exhibit a stronger $1/n^2$ scaling.

%%%%%%%%%%%%%%%%%%%%%%%%%%%%%%%%%%%%%%%%%%%%%%%%%%%%%%%%%%%%%%%%%%%
\end{document}